\documentclass[twocolumn,amsmath,floatfix,prb]{revtex4}
\usepackage{graphicx,bm}
\begin{document}

\title{Free Energy Landscape of Protein-like Chains with 
Discontinuous Potentials}

\author{Hanif Bayat Movahed$^a$}\email{hbayat@chem.utoronto.ca}
\author{Ramses van Zon$^{a,b}$}\email{rzon@scinet.utoronto.ca}
\author{Jeremy Schofield$^a$}\email{jmschofi@chem.utoronto.ca}

\affiliation{$^a$Chemical Physics Theory Group, Department of Chemistry, University of Toronto, 80 St.\ George Street, Toronto, Ontario M5S 3H6, Canada\\
$^b$SciNet High Performance Computing Consortium, University of Toronto, 256 McCaul Street, Toronto, Ontario M5T 1W5, Canada
}

\date{\today}

\begin{abstract}
In this article 
the configurational space of  two 
simple protein models consisting of polymers composed of a periodic sequence of four different kinds of monomers is studied as a function of temperature. 
 In the protein models,
hydrogen bond interactions, 
electrostatic repulsion, and covalent bond vibrations are modeled by discontinuous step, shoulder and square-well potentials, respectively. 
The protein-like chains exhibit a secondary alpha helix structure in their folded states at low temperatures, and allow a natural definition of a configuration by considering which beads are bonded. Free energies and entropies of configurations are computed using the parallel tempering method in combination with hybrid
Monte Carlo sampling of the canonical ensemble of the 
discontinuous potential system.
The probability of observing the most common configuration is used to 
analyze the nature of the free energy landscape, and it is found that
the model with the least number of possible bonds
exhibits a funnel-like free energy landscape at low enough temperature for chains with fewer than 30 beads. For longer proteins, the landscape consists of several minima, where the configuration with the lowest free energy changes significantly by lowering the temperature and the probability of observing the most common configuration never approaches one due to the degeneracy of the lowest accessible potential energy.

\end{abstract}

\maketitle

\section{Introduction}
\label{sec:Intro}

Statistical mechanical modeling has helped significantly in addressing 
the question of why protein folding occurs so rapidly in spite of the
astronomically large number of possible configurations available to a protein.
It has been suggested that folding occurs on funnel-shaped energy landscapes rather than involving a single microscopic pathway through a complicated landscape\cite{Dill:64}. Onuchic, Dill, Wolynes and co-workers proposed that a ``folding funnel'' is the special characteristic of foldable proteins that directs the folding protein into the native state without the need for a definite pathway\cite{Onuchic:65, Dill:62, Wolynes:45, Thirumulai:67, Socci:20, Wolynes:21, Dill:66, Wolynes:22}.  According to this picture,
topological features of the free energy landscape, defined in a coarse-grained sense by averaging over 
conformations of the protein with similar characteristics, assist the folding process by channeling or funneling the evolution of configurations. The folding of a protein is viewed as a process in which the protein glides down in the funnel-shaped free-energy landscape as the temperature
drops or as time progresses along a multitude of different paths towards its native structure\cite{Gin:19,Socci:20,Wolynes:21,Wolynes:22}.  According to this viewpoint, structures with low free energies are situated within a basin of a broad energy valley and a protein in a configuration associated with one of the valleys can move quickly in
the funnel to the lowest free energy state. 

Of course the true free energy landscape is never a simple funnel, and the configurational space 
of a protein is a highly multi-dimensional space. Even for small proteins, its dimensionality ranks in the several hundreds\cite{McLeish:24}. Within this high dimensional space, the free energy landscape can feature many local minima separated by energetic and entropic barriers. 

Although the free energy landscape of proteins 
is often considered to be a key component in
understanding the mechanisms of protein folding,
the characterization of the structure of the free
energy landscape is nebulous
due to the difficulty of identifying the relationship between different
conformations of proteins and determining whether
particular configurations are within the same configurational basin.  
The difficulty of identifying conformations  of proteins is 
compounded by the computational challenge of achieving converged sampling
of available configurations  for realistic protein models.

In this paper, studies of the energy landscape of a protein-like chain in the absence of any fluid are presented. 
Such a study is not feasible at present for realistic models of proteins, so simplified models are used
to capture the basic behavior of proteins.
Discontinuous potentials are used for the interaction potentials, where attraction and repulsion are defined as step and shoulder potentials respectively. The Hybrid Monte Carlo (HMC) method\cite{Duane:30} is applied for the sampling of the energy landscape of a protein-like chain in which the Monte Carlo sampling is done using parallel tempering (PT) and the generation of trial configurations is carried out by discontinuous molecular dynamics (DMD). The PT method \cite{Swendsen:6, Greyer:7, whittington:18} improves the convergence properties of Monte Carlo sampling by decreasing the correlation length
of samples in the Markov chain of states\cite{Earl:8}.  

It is shown that for two
simple protein models, each consisting of a periodic sequence of four different kinds of bead, the folded state exhibits a secondary alpha helix structure. It is demonstrated that the relative configurational entropies of the protein-like chains are independent of temperature for
the discontinuous potential models, which makes it possible to compute the relative configurational entropies and the free energies of the configurations very accurately. Relative configurational free energies at different temperatures can be determined from relative populations at those temperatures. The free energy results can be interpreted in terms of the free energy landscape picture.
Such understanding of the free energy landscape is the main objective of this work.

In Sec.~\ref{Pr-model} the models and their parameters are described, and it is shown that relative configurational entropies are temperature independent in the models. A simplified three state model is also presented to facilitate the interpretation of the simulation results.
In Sec.~\ref{sec:Results}, the results for the observed structures, configurational entropy and free energy differences are presented, and the shape of the free energy landscape is analyzed both
for short and long chains. Conclusions are given in Sec.~\ref{sec:concl}.

\section{Models of the protein-like chain}
\label{Pr-model}

In this article we consider a \emph{beads on a string} model of a protein-like chain in which each bead represents an amino acid or residue. The chain consists of a repeated sequence of four different kinds of beads. While having four different types of beads is not enough to represent the twenty different types of amino acids, it preserves at least some of the differences between amino acids. The interactions between these beads are designed to mimic the interactions that lead to the formation of common motifs in protein structure, such as the alpha helix. Previous studies suggest that chains containing only $6$, $8$ or $12$ monomers are too short to fold into compact states at low temperatures, while somewhat longer chains with $25$ monomers can capture folded helical states\cite{Athawale:13}. Here, chains of moderate lengths of $15$ to $35$ beads have been used to facilitate the exploration of the free energy landscape.

The models analyzed here allow for attractive interactions, intended to mimic hydrogen bonds between non-adjacent residues, between beads separated from each other by $4n$ beads, where $n\geq1$, and with additional restrictions on the possible hydrogen bonds to be specified below. Several versions of the models of protein-like chains have been considered, but only the results for two of them are presented here. Models were selected based on the  similarity of preferred structures in the model to those observed  in real proteins. 

To make contact with real proteins, physical units are used in the definition of the model, although these should not be taken too literally.   In particular, lengths will be expressed in \AA ngstr\"oms, energies in kJ/mol and masses in atomic mass units.

\begin{figure}[t]
  \centering
  (a)~~~~~~~~~~~~~~~~~~~~~~~~~~~~~~~~~~~~~~~(b) \\
  \includegraphics[width=0.45\columnwidth]{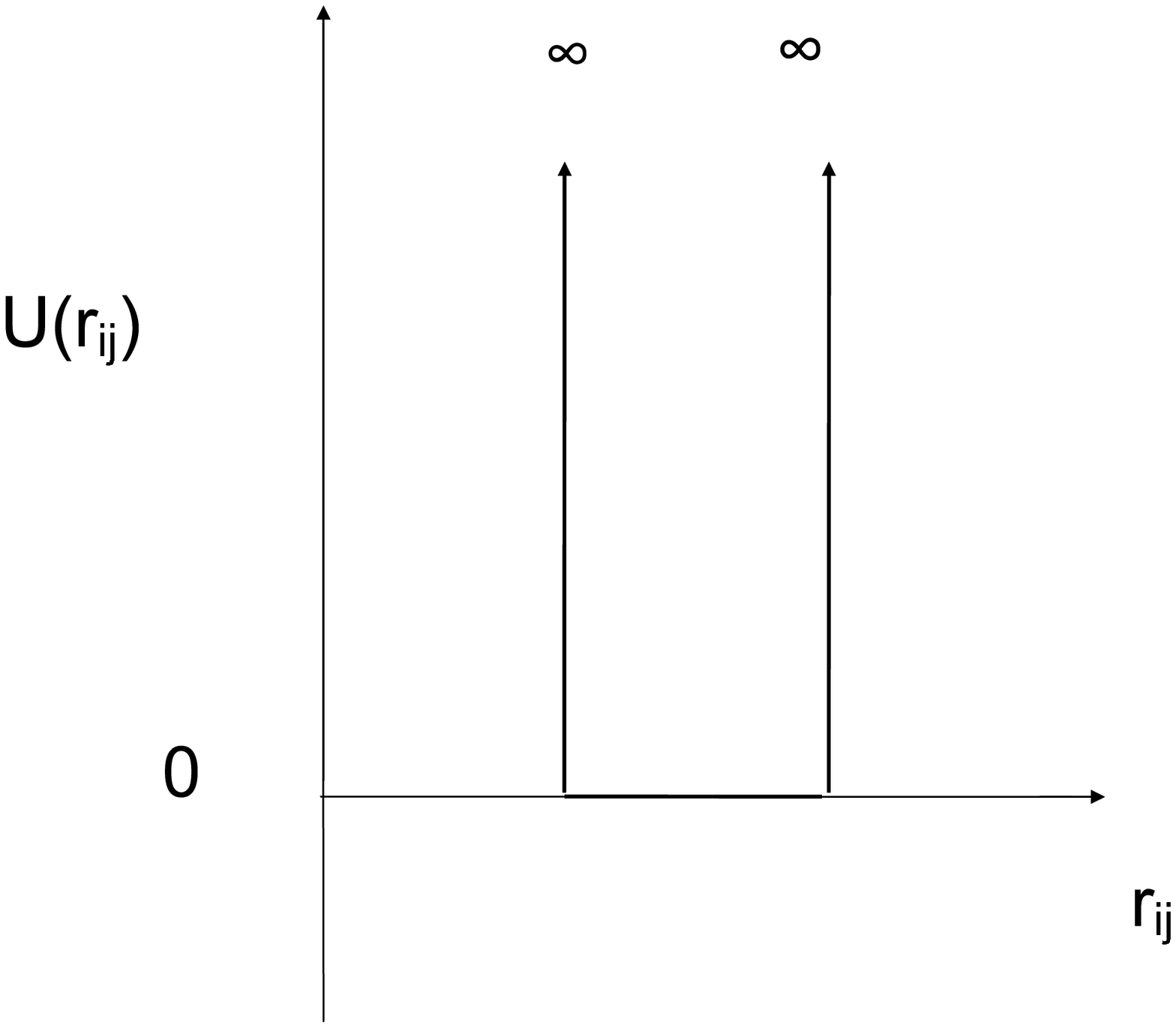}
  \includegraphics[width=0.45\columnwidth]{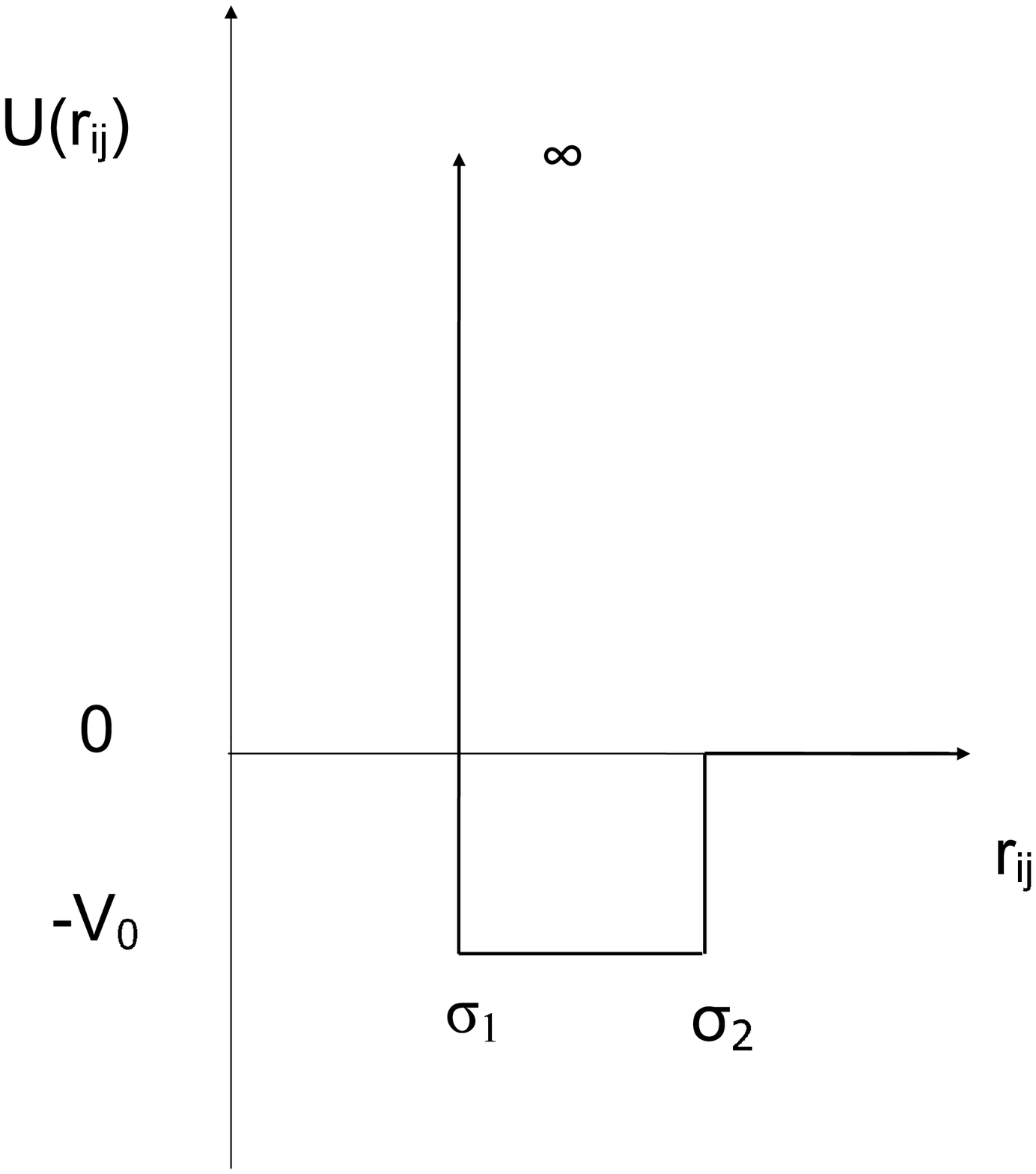}\\
  (c)~~~~~~~~~~~~~~~~~~~~~~~~~~~~~~~~~~~~~~~(d) \\
  \includegraphics[width=0.45\columnwidth]{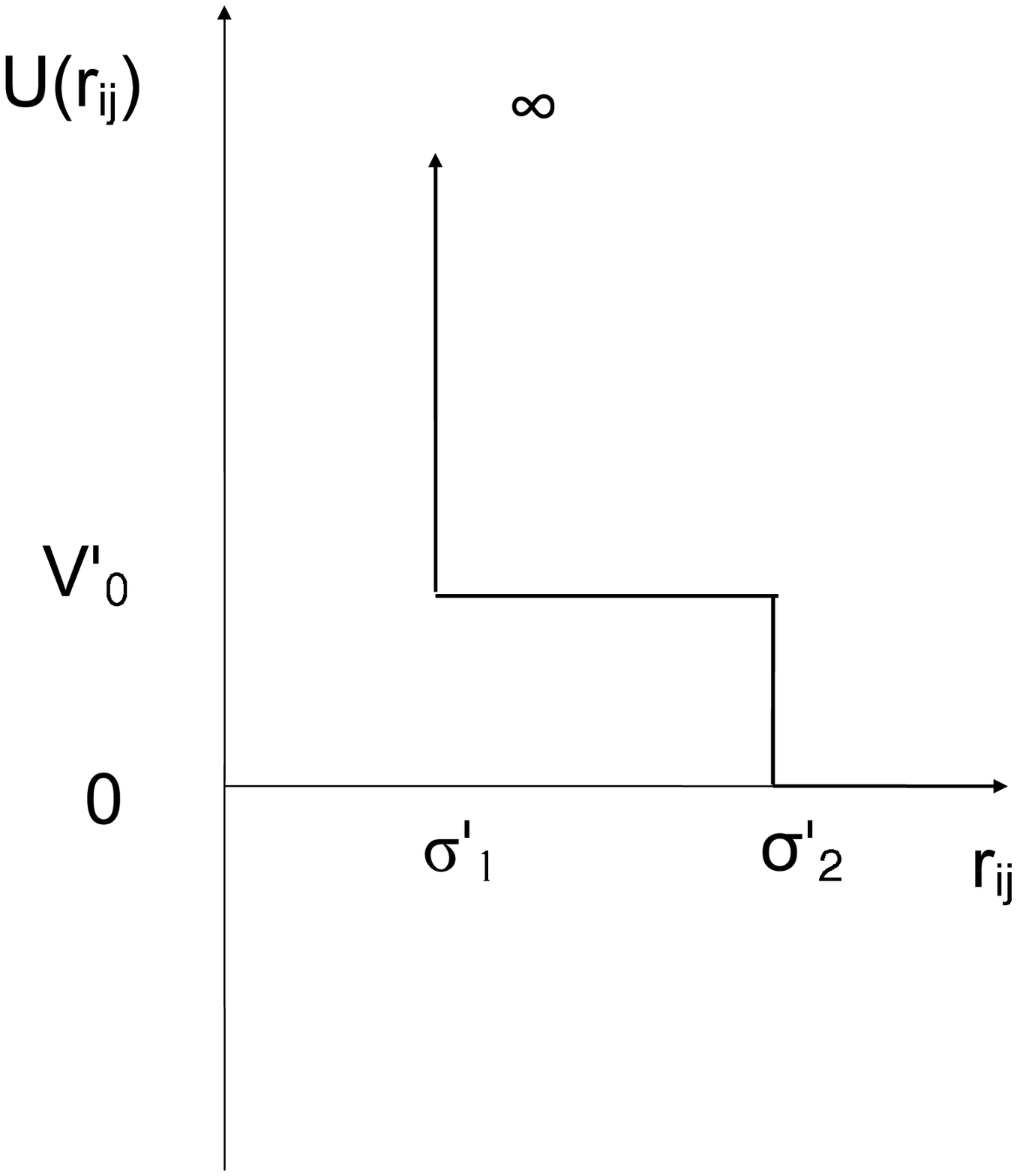}
  \includegraphics[width=0.45\columnwidth]{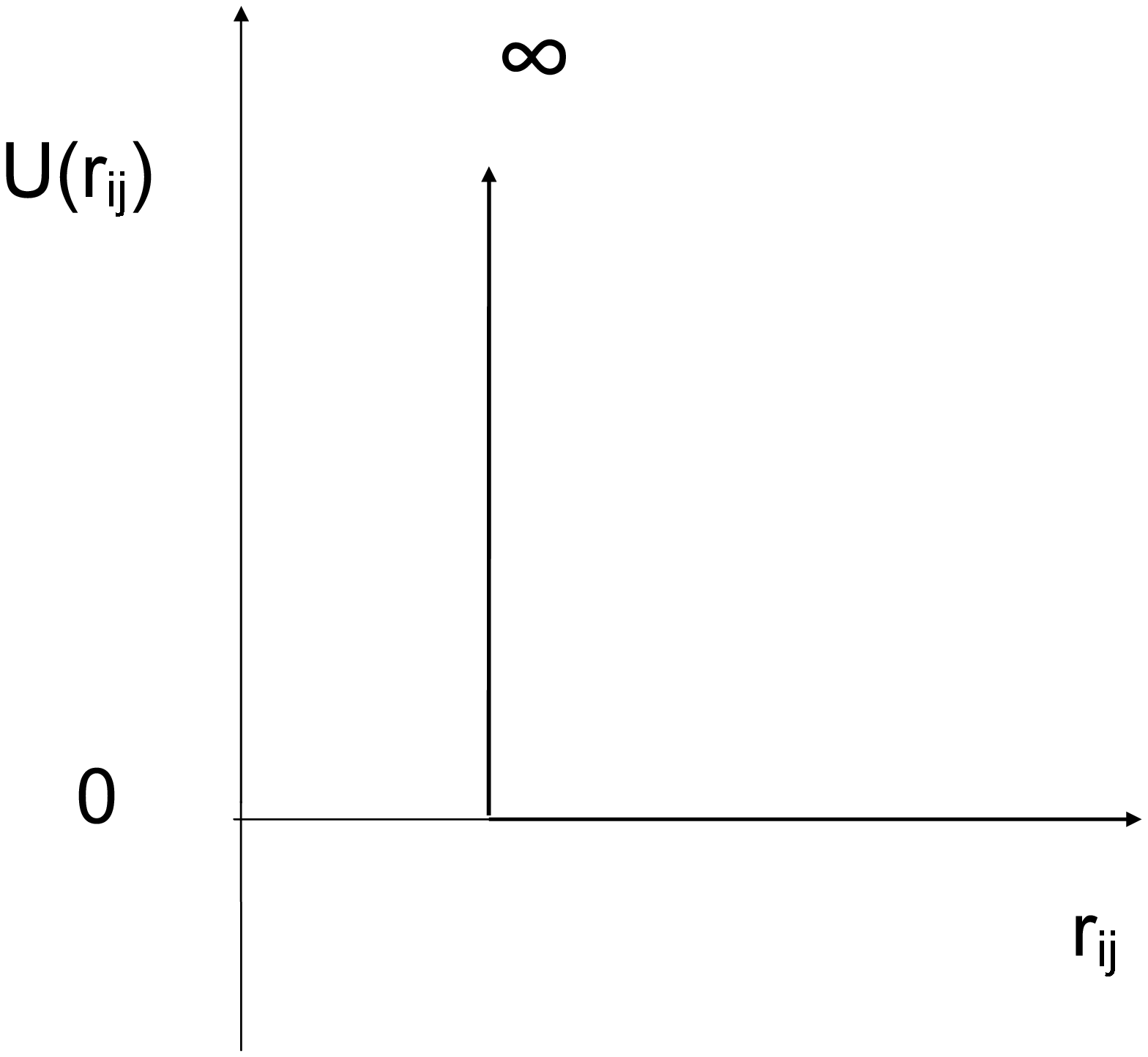}
  \caption{Model~potentials: the (a) infinite square-well potential,
    (b) attractive step potential, (c) repulsive shoulder potentials,
    and (d) hard core repulsion.}
  \label{fig:modelpot}
\end{figure}

The two models analyzed here  differ in the hydrogen-bond potentials, while other interactions are the same. In total, four different potentials are used in these models.  The first kind of potential acts between the nearest and the next nearest neighbors and restricts the distance between the beads to specific ranges by applying an infinite square-well potential similar to Bellemans' bonds model\cite{Bellemans:26}.  Fig.~\ref{fig:modelpot}(a) shows the shape of this kind of potential. To mimic a covalent bond between two consecutive amino acids in the protein, the distance between two neighboring beads is restricted to the range 3.84~\AA\ to 4.48~\AA. This potential allows these distances to vibrate around values close to the distance between stereocenters used in Ref.~\onlinecite{ZHOU:15}. Bond angle vibrations are similarly represented by defining infinite
square-well potentials between next-nearest neighbors in the chain. Restricting their distance to a range from 5.44~\AA\ to 6.40~\AA\ generates a vibration angle between 75$^\circ$ and 112$^\circ$. For simplicity, dihedral angles are not considered in our models, but as discussed later, some restrictions on hydrogen bonds are employed to create rigidity in the backbone of the protein-like chain similar to the rigidity that results from the dihedral angle interactions in more detailed potentials.

Hydrogen bonds are modeled by an attractive square-well potential, depicted in Fig.~\ref{fig:modelpot}(b). In all models investigated here, the attractive forces are defined between beads $i$ and $i+4n$ (with $n$ integer) to resemble the hydrogen bonds in alpha helix structures. However, the two models differ in the possibility of these attractive bonds and the values of $i$ and $n$.

In the first model, named model A, the attractive interactions act between half the same type beads such that bonds can be formed between two beads both with an index of the form $i=4k+1$, or both with an index of the form $i=4k+3$, where $k$ is an integer number.

\begin{figure*}
  \centering
(a)\hspace{7.5cm}(b)\\
  \includegraphics[width=0.7\columnwidth]{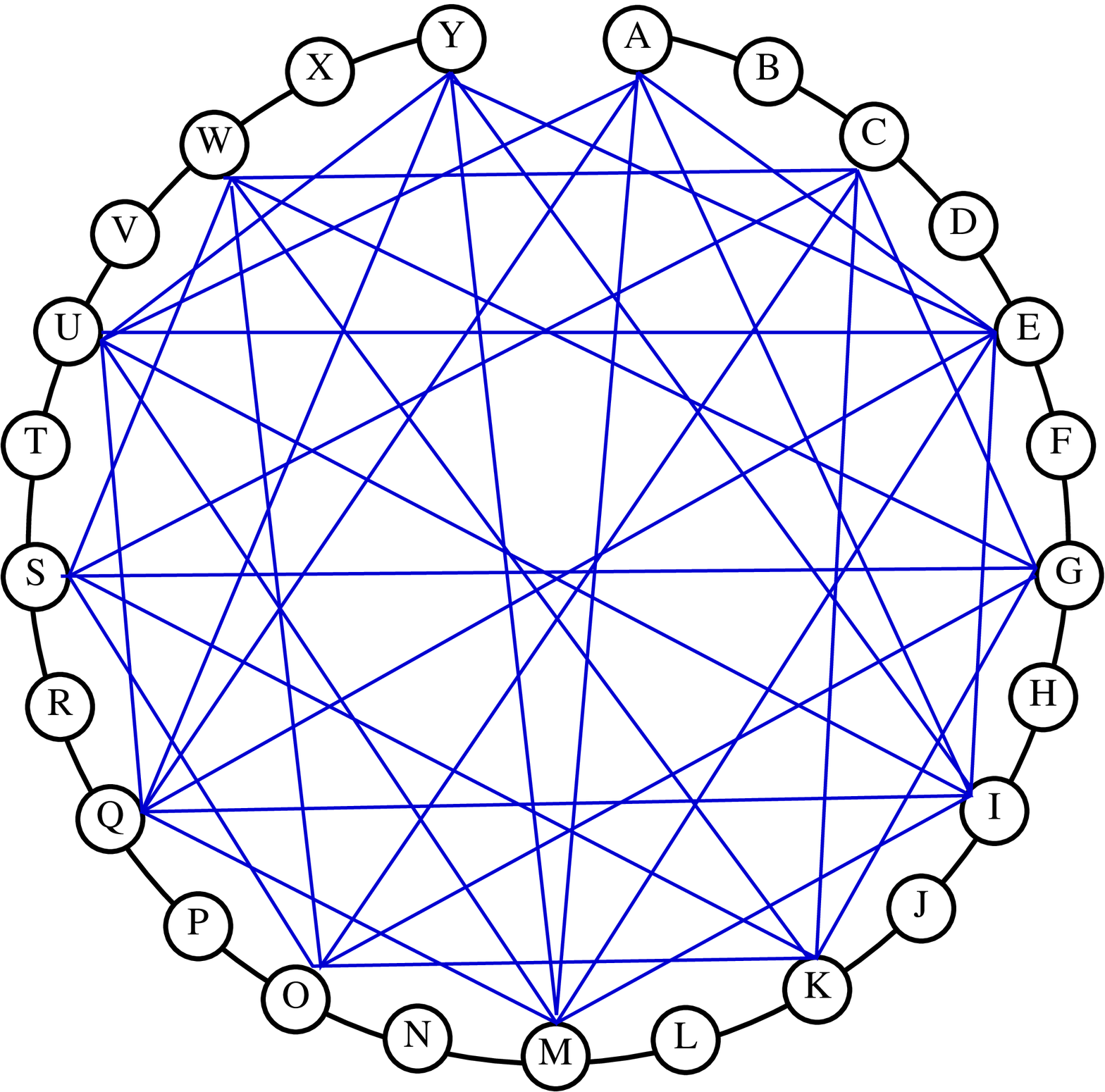}~~~~~~~~~
  \includegraphics[width=0.7\columnwidth]{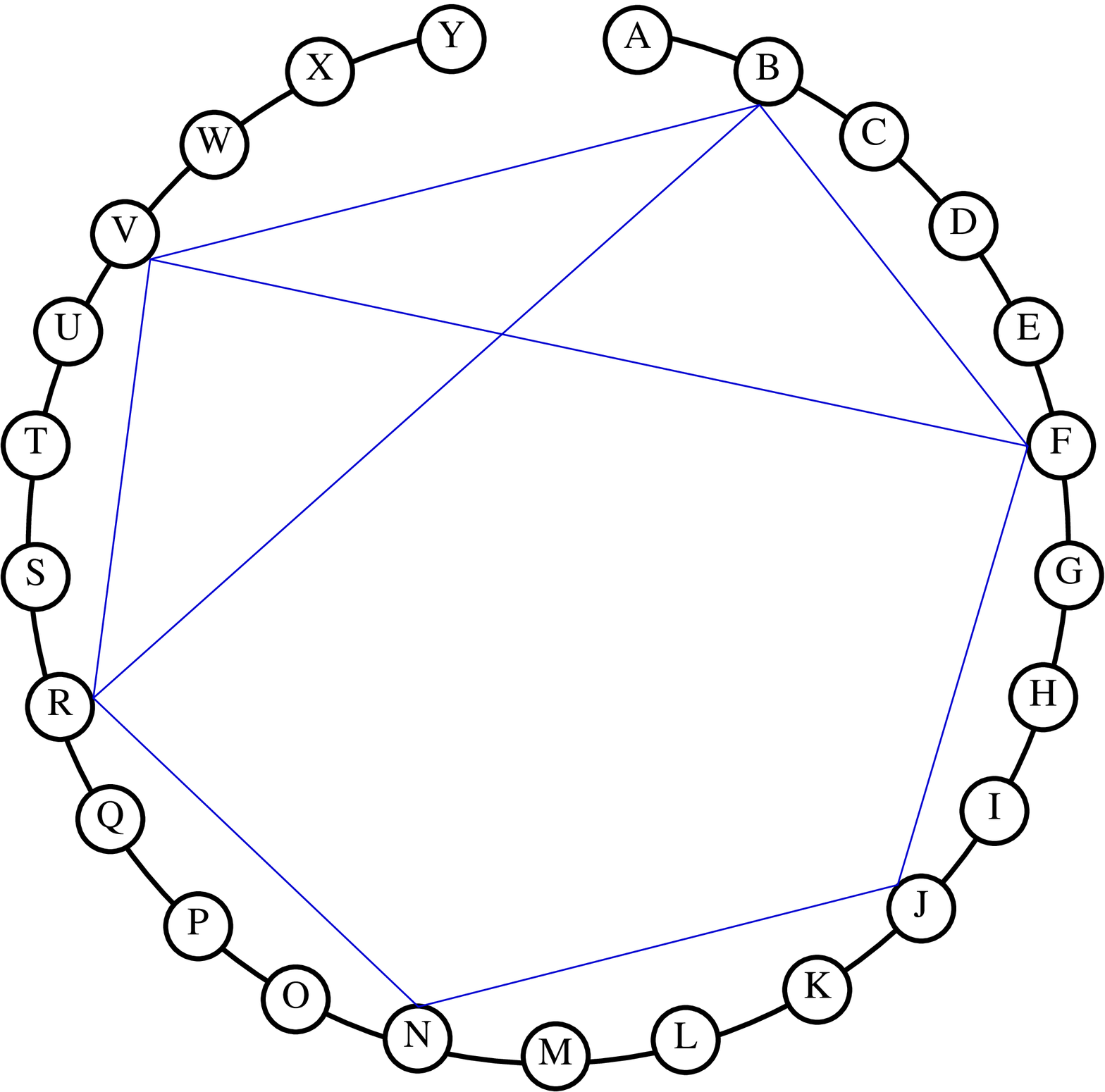}
  \caption{Possible attractive bonds of (a) model~A, and (b) model~B for a chain of 25 beads.}
  \label{fig:interactions}
\end{figure*}

In the second model, model~B, only the beads with index $i=4k+2$ can bond with each other, and $n$ cannot be $2$ or $3$. This means that there is no attractive bond between beads separated along the chain by eight or twelve beads. Bonds between beads $i$ and $i+8$ as well as $i$ and $i+12$ are disallowed to make the occurrence of turns more difficult in the protein-like chain and effectively make it more rigid. This restriction has a similar function to torsional interaction potentials defined in terms of dihedral angles along the backbone of the chain in more detailed models where they prevent a protein from bending over easily.
In Fig.~\ref{fig:interactions} the possible attractive bonds for the two models are presented for a chain of length 25 in which subsequent beads were labeled A through Y. It will be shown that the two models have different thermodynamic characteristics and important qualitative differences in their free energy landscape due to the difference in the hydrogen bonding interactions.

For both models, the parameters for the attractive square-well potential, $\sigma_1$ and $\sigma_2$, are chosen to be 4.64~\AA\ and 5.76~\AA\, with a mid point of 5.2~\AA, which is close to the translation of 5.4~\AA\ along each turn of an alpha helix. Compared to covalent bonds, these attractive interactions act across longer distances. The 
depth of the potential well 
$\epsilon$ is  20~kJ/mol and the mass of each bead is set to $2\times10^{-25}$kg,\! which is close to 120 atomic mass~units.

To represent electrostatic interactions of the atoms, repulsive interactions act between beads $1+4k$ and $4k'$, where $k$ and $k'$ are integers and $k \neq k'$. The repulsive interaction takes the form of a shoulder potential, shown in Fig.~\ref{fig:modelpot}(c). The range of the shoulder is set to be from 4.64~\AA\ to 7.36~\AA, while the height is $0.9 \epsilon$.
The effect of changing the number of step repulsions in a few models was evaluated in terms of minimizing the free energy. It turned out that changing the number of repulsions does not have a huge impact on the shape of free energy landscape around the native structure point. Since the repulsion between the beads increases the potential energy while decreasing the configurational entropy, the most common structures at low temperatures do not have any repulsive interactions. Therefore, the two models differ only in their attractive potentials, while their repulsive interactions are the same.

Finally, all other bead pairs for which no covalent bonds, hydrogen bonds or shoulder repulsive interactions are defined interact via a hard sphere repulsion to account for excluded volume interactions at short distances, depicted in Fig.~\ref{fig:modelpot}(d). The hard sphere diameter is set to be 4.64~\AA, which is slightly different from the value of 4.27~\AA\ used by Zhou et al.\cite{ZHOU:15}.

The reduced temperature is defined as $T^*=(k_bT)/\epsilon$, where $\epsilon$ is the potential depth of the square-well attractive interactions, and $\beta^*$ is the inverse of the reduced temperature, $\beta^*=1 /T^*$. Given the value of $\epsilon=$~20kJ/mol, $T^*=1.0$ corresponds to $2400$ K. This means that $\beta^*=8$ ($T^*=\frac{1}{8}$) roughly corresponds to standard room temperature, $300$ K.

\subsection{Definition of configurations}
\label{sec:Bondrepresentation}

One of the advantages of using discontinuous potentials is the ease of comparing configurations. The bonds are defined using the specific range of bead separations $r_{ij}$ in which the potential energy $V=\frac{1}{2}\sum_{ij}U(r_{ij})$ is equal to a specific, non-zero value. Since only one 
attractive bond can exist between each bead pair $(i,j)$ in the current models, each configuration or structure can be represented by a matrix of interactions in which the entry at row $i$ and column $j$ is unity if $i$ and $j$ are bonded and zero otherwise.
Because bonded interactions largely determine the form of the protein,  this matrix can be used to identify the configuration of the protein-like chain. Thus, by comparing the matrices, identical structures can be easily found.

However, for ease of presentation, a more readable alphabetical notation for configurations is applied. Each bead is represented by a subsequent letter from the alphabet and each bonded interaction is shown by a pair of letters. The two dimensional matrix can thus be represented by a string of alphabetical pairs. Since most of the studied cases involve 25-bead chains, A to Y have been used to label different beads. For chains longer than 26 beads, both capital and small letters can be used.

\subsection{Temperature independence of relative configurational entropies}
\label{sec:Entropies}

The definition of configurations presented above was based on the presence of attractive bond interactions.
Within the model, having a certain set of bonds (and no others) leads to a specific potential energy $U_c$ for each configuration $c$.
As shown below, this leads to a temperature independent relative configurational entropy.

The configurational entropy of any particular configuration $c$ is the entropy of a sub-ensemble in which the phase points are restricted to those of configuration $c$.
The discrete nature of the interactions
allows configurational space to be partitioned into microstates
by defining an index function for a configuration $c$ that depends on the set of spatial
coordinates of the chain $\bm{R}$
\begin{eqnarray*}
\chi_c( \bm{R} ) = \left\{ 
\begin{array}{ll}
1 & \mbox{if only bonds in $c$ are present,} \\
0 & \mbox{otherwise}.
\end{array}
\right.
\end{eqnarray*}
The partitioning of configurational space arises naturally by
expanding the product in the indentity
\begin{eqnarray}
1 &=& \prod_{i=1}^{n_b} \left( 1 - H(x_i - \sigma_2) + H(x_i- \sigma_2)\right) \nonumber \\
&=&
\prod_{i=1}^{n_b} 
\left( H(\sigma_2 - x_i) + H(x_i- \sigma_2)\right) \\
 &=& \sum_{k=1}^{n_s} \chi_{c_k} (\bm{R}) , 
\label{partition}
\end{eqnarray}
where $n_b$ is the number of attractive bonds in the model, $n_s =
2^{n_b}$ is the number of microstates, 
and $H(x)$ is the Heaviside function
\begin{eqnarray*}
H( x ) = \left\{ 
\begin{array}{ll}
1 & x \geq 0 \\
0 & \mbox{otherwise}.
\end{array}
\right.
\end{eqnarray*}
In Eq.~(\ref{partition}), $x_i$ is the distance between monomers in
the $i$th bond, and $\sigma_2$ is the critical distance at which an 
attractive hydrogen bond is
formed.  For notational simplicity, we order the index of
configurations based on the number of bonds starting with the
configuration with no bonds, $\chi_1(\bm{R} ) = \prod_{i=1}^{n_b} H (x_i
- \sigma_2)$, and ending with the configuration with the maximum number of
bonds, $\chi_{n_s}(\bm{R} ) = \prod_{i=1}^{n_b} H(\sigma_2-x_i)$.

In the canonical ensemble, the probability $f_{\rm obs}(c,T)$ of observing a configuration $c$ at temperature $T$ is
\begin{equation}
  f_{{\rm obs},c} = e^{-\beta (F_c-F)},
  \label{eq:fobs}
\end{equation}
where $F_c$ is the free energy of configuration C, and $F$ is the full free energy of the system.
By definition, one has
\begin{equation}
  e^{-\beta F_c} = \frac{1}{h^{3N}}\int\! dR\, dP\,
  \chi_c(R) e^{-\beta \left[\sum_{i=1}^N
      \frac{|p_i|^2}{2m}+V(R)\right]}
\label{eq:Fc},
\end{equation}
where $N$ is the number of beads, $m$ is their mass, and $V$ is the potential energy function.
The configurational entropy is related to $F_c$ via
\begin{equation}
  F_c = E_c - T S_c,
\label{eq:Sc}
\end{equation}
where $E_c$ is the average energy of configuration $c$ at temperature $T$.
Since its potential energy $V$ is always equal to $U_c$ when it
is finite and $\chi_c=1$, one has
\begin{equation}
  E_c=U_c+\frac{3}{2}Nk_BT.
\label{eq:Uc}
\end{equation}
Combining Eqs.~\eqref{eq:Fc}-\eqref{eq:Uc}, one finds
\begin{equation}
\label{eq:entropy}
  S_c = \frac32Nk_B\ln\left(\frac{2\pi me}{\beta h^2}\right) +
  k_B\ln\int^\prime dR\,\chi_c(R),
\end{equation}
where the integral is restricted to sum over configurations that satisfy all
geometric constraints due to the infinite square-well and hard core repulsions.
Thus the \emph{relative} entropy of two configurations $c_1$ and $c_2$ at a specific temperature is
\begin{equation}
  \Delta S_{c_1c_2}
  = S_{c_1} - S_{c_2}
  = k_B \ln \frac{\int^\prime dR\,\chi_{c_1}(R)}{\int^\prime dR\,\chi_{c_2}(R)},
\end{equation}
which does not depend on temperature.

From Eqs.~\eqref{eq:Uc} and ~\eqref{eq:entropy} it can be concluded that the free energy of a configuration is
\begin{equation}
\label{eq:freeE}
  F_c = U_c-\frac32Nk_BT\ln\left(\frac{2\pi m}{\beta h^2}\right)- k_BT\ln\int^\prime\,  dR\,\chi_c(R),
\end{equation}
where the second term is the same for all the configurations at temperature $T$.

Because relative configurational entropies do not depend on temperature, relative entropies can be determined from a single run at a temperature $T$, using
\begin{eqnarray}
  \Delta S_{c_1c_2} &=& \frac{\Delta E_{c_1c_2}-\Delta F_{c_1c_2}}{T}
  \nonumber\\
  &=& \frac{\Delta E_{c_1c_2}}{T}+k_B\ln\frac{f_{\rm obs}(c_1,T)}{f_{\rm obs}(c_2,T)}\\
  &=& \frac{\Delta U_{c_1c_2}}{T}+k_B\ln\frac{f_{\rm obs}(c_1,T)}{f_{\rm obs}(c_2,T)}.
\label{eq:entropyII}
\end{eqnarray}

Therefore, no approximation is necessary to calculate the relative configurational entropies in contrast to molecular dynamics (MD) studies utilizing smooth potentials (see e.g.\ Ref.~\onlinecite{Wei:31}).

\subsection{Simulation Techniques}
\label{sec:Simulation-Techniques}

The simulation results presented here were obtained utilizing a sampling method that uses a combination of dynamical updates based on DMD and PT
exchange moves.
In this approach, a number of replicas are updated simultaneously using
molecular dynamics (appropriate for the discontinuous potential systems\cite{Rappa:4}) for
a fixed amount of time.  At the start of a dynamical update, the velocities of all
beads in the chain are drawn from the Maxwell-Boltzmann distribution for 
each replica at the temperature appropriate for that replica.  Since the DMD
is time-reversible, exactly conserves energy and preserves 
phase space volume\cite{Hernandez:9},
the limit distribution of the Markov chain of states for each replica is
canonical at the temperature of the Markov chain\cite{Duane:30}.  Furthermore
since the total energy is conserved exactly in the dynamics, the updates provide a rejection-free means of moving all degrees of freedom simultaneously.  To enhance
the sampling efficiency, the dynamical updates are combined with replica
exchange updates.  The replica exchange moves are
designed so that the states at each temperature are canonically distributed\cite{Swendsen:6,Greyer:7}.
The process of drawing velocities, DMD dynamics, and PT exchange moves is repeated until enough independent statistics on the frequency at which different configurations are seen is gathered.

\subsection{Simplified three state model}
\label{sec:threestate}
\begin{figure}[b]
  \centering
  \includegraphics[width=\columnwidth]{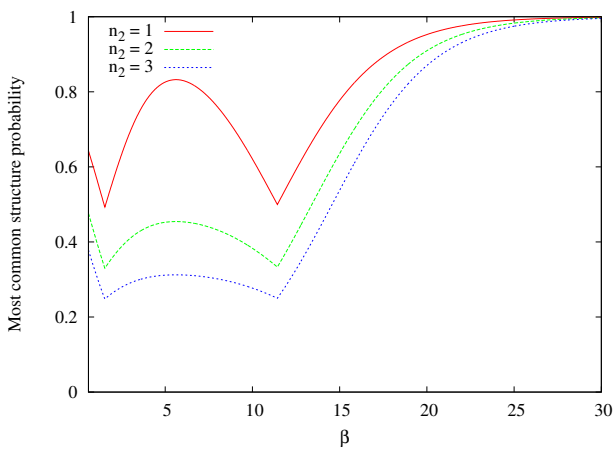}
  \caption{Variation of the probabilities of the most common
structure versus the inverse temperature  
for the three state model with parameters  $E_2=-0.65$, $S_2=4$, $S_3=5$, and three different values for $n_2$. 
\label{fig:threestate}}
\end{figure}

In the simulations, one can easily measure the frequency of occurrence
$f_{\rm obs}(c)$ of configurations $c$ at each temperature in the PT
replica set. The accuracy of $f_{\rm obs}(c)$ is $\mathcal
O(\sqrt{f_{\rm obs}(c)})$, and thus is highest for the most frequently
occurring (dominant) structure.  For that reason, below, we will often
plot the observed frequency $f^*$ of the most common structure, i.e. $f^*=\max_c
f_{\rm obs}(c)$, as a function of the inverse temperature $\beta$.  To
facilitate the interpretation of such a plot, it is helpful to
consider its form for the following, simplified three-state model. The
three states are configurations with energies $E_1$, $E_2$ and $E_3$,
and entropies $S_1$, $S_2$, and $S_3$, respectively.  
As in the actual
model of the protein-like chain, the values of entropies do not depend
on temperature. The second state will furthermore be taken to be $n$
fold degenerate (this is thus really a $n+2$ state model). 

For each state in this model, the observational frequency is
\begin{eqnarray}
  f_{\rm obs}(c,\beta) = \frac{e^{-\beta E_c + S_c/k_B}}{Z(\beta)},
\end{eqnarray}
where $c$ ranges from 1 to 3 and
\begin{eqnarray}
  Z(\beta) =  e^{-\beta E_1 + S_1/k_B} + n e^{-\beta E_2 + S_2/k_B} +  e^{-\beta E_3 + S_3/k_B}.
\end{eqnarray}

We will assume that  $E_1<E_2<E_3$ and $S_1<S_2<S_3$, such that
configuration 1 models the native state with lowest energy and lowest
entropy, configuration 3 models the unfolded state with high energy
and high entropy, while configuration(s) 2 can be interpreted as
intermediate.  Because only relative energies and entropies affect
$f^*$, we can set $E_3$ and $S_1$ to zero. Furthermore, one
can fix the temperature scale by setting $E_1$ to $-1$. That leaves
just four parameters in the model: $n_2$, $E_2$, $S_2$ and $S_3$
(subject to the constraint that $S_2<S_3$).

Figure~\ref{fig:threestate} shows three examples of the behavior of
$f^*$ for this model, corresponding to the following choices of the
parameters: $E_2=-0.65$, $S_2=4$, $S_3=5$, and $n_2=1$, $2$, and $3$,
respectively.  One sees a 'bouncing' signal as subsequent states
become dominant when temperature is varied.  There are cusp-shaped
minima where the identity of the dominant state changes. At that
point, several configurations are equally likely. If one neglects the
other, non-dominant, configurations at that point, then the value of
$f^*$ at a cusp should be one over the number of competing structures,
and this is borne out by the plots shown in Fig.~\ref{fig:threestate},
which show cusp depths close to 1/2, 1/3 and 1/4, respectively. As
$\beta$ increases (temperature gets lower), the frequency of observing
state 1 (the `native' state) reaching almost 100\%.

One can expect similar results for the protein-like chain model used
in the simulations. The main difference with the three-state model is
the presence of many more configurations.  Some of these extra states with be irrelevant (have negligible $f_{\rm obs}$) because of their low entropy, but one could expect to see extra bounces in the plots for the real model from some addition relevant states.

\section{Results}
\label{sec:Results}

\begin{table*} 
  \begin{tabular} {p{0.6cm}|p{14.52cm}|p{1.4cm}}
    $\beta^*$ & the most common structure &
    $f_{\rm obs}$(\%)  \\
    \hline
    1.5& No bond &14.2$\pm$0.6 \\
    14.0& AE AI AY CG CK CS CW EI GK GO GS IY KO KS KW MQ MU OS QU SW &9.7$\pm$0.6\\
    24.0& AE AI AY CG CK CS CW EI GK GO GS IY KO KS KW MQ MU OS QU SW &10.6$\pm$0.6 \\
    38.4& AQ AU AY CG CO CS CW EI EM GK GO GS GW IM KO KS OS QU SW UY &8.5$\pm$0.6\\
    57.5& AQ AU AY CG CO CS CW EI EM GK GO GS GW IM KO KS OS QU SW UY &7.7$\pm$0.6\\
    72.5&  AE AI AM CG CK CO CS EI GK GS GW KO KS KW OS OW QU QY SW UY &8.1$\pm$0.6\\
    87.5&  AE AI AM AQ AU AY CG CK EI EY GK IM IQ IY MQ MU OS QU QY SW UY  &8.2$\pm$0.6\\
    \hline
    $\beta^*$ & the second most common structure & $f_{\rm obs}$(\%)  \\
    \hline
    1.5 & SW & 2.1$\pm$0.2 \\
    14.0 & AE AI AY CG CK CO CS CW EI GK GO IY KO KW MQ MU OS OW QU SW & 8.7$\pm$0.6\\
    24.0 & AE AI AY CG CK CO CS CW EI GK GO IY KO KW MQ MU OS OW QU SW & 9.6$\pm$0.6 \\
    38.4 & AE AI AY CG CK CO CS CW EI GK GO IY KO KW MQ MU OS OW QU SW & 6.6$\pm$0.4\\
    57.5 & AE AI AM CG CK CO CS EI GK GS GW KO KS KW OS OW QU QY SW UY & 4.5$\pm$0.4\\
    72.5 & AQ AU AY CG CO CS CW EI EM GK GO GS GW IM KO KS OS QU SW UY & 7.5$\pm$0.4\\
    87.5 &  AE AU AY CG CS CW EY GK GS GW IM IQ KO KS KW MQ OS OW SW UY & 6.6$\pm$0.6
  \end{tabular}
  \caption{Most common configurations of the model~A 25-bead chain.}
  \label{tab:Table170A}
\end{table*}

\begin{table} 
  \begin{tabular}{r|l|l}
    $\beta^*$ & the most common structure &
    $f_{\rm obs}$(\%)  \\\hline
    1.5 & No bond & 22.4 $\pm$ 1.2  \\
    3.0 & No bond & 6.7 $\pm$ 1.0   \\
    3.5 & BF JN &  4.0 $\pm$ 0.6 \\
    4.2 & BF FJ NR RV & 6.5 $\pm$ 0.8 \\
    4.5 & BF BR BV FJ FV JN NR RV & 7.5 $\pm$ 1.0  \\
    5.3 & BF BR BV FJ FV JN NR RV & 46.4  $\pm$ 1.6 \\
    6.0 & BF BR BV FJ FV JN NR RV &  76.0 $\pm$ 1.2 \\
    7.5 & BF BR BV FJ FV JN NR RV  & 94.1 $\pm$ 0.8 \\
    13.5 & BF BR BV FJ FV JN NR RV & 99.9 $\pm$ 0.0 \\
    \hline
    $\beta^*$ & the second most common  &$f_{\rm obs}$(\%) \\
    \hline
    1.5 & BF & 3.5 $\pm$ 0.6  \\
    3.0& BF & 5.6 $\pm$ 0.8\\
    3.5 & BF NR & 4.0 $\pm$ 0.6 \\
    4.2 & BF FJ JN RV & 4.9 $\pm$ 0.8\\
    4.5 & BF FJ JN NR RV  & 6.4 $\pm$ 0.8\\
    5.3 &  BF BR BV FJ JN NR RV &  10.1 $\pm$ 0.8\\
    6.0 & BF BR BV FJ JN NR RV & 6.8 $\pm$ 0.8 \\
    7.5 & BF BR BV FJ JN NR RV & 1.9 $\pm$ 0.6\\
    13.5 & N/A & N/A
  \end{tabular}
  \caption{Most common configurations of the model~B 25-bead chain.}
  \label{tab:modelB-dominant}
\end{table}

\subsection{Free energy landscape}
\label{sec:subsecenergyland}

\begin{figure}[t] 
  \centering
  \includegraphics[width=0.8\columnwidth]{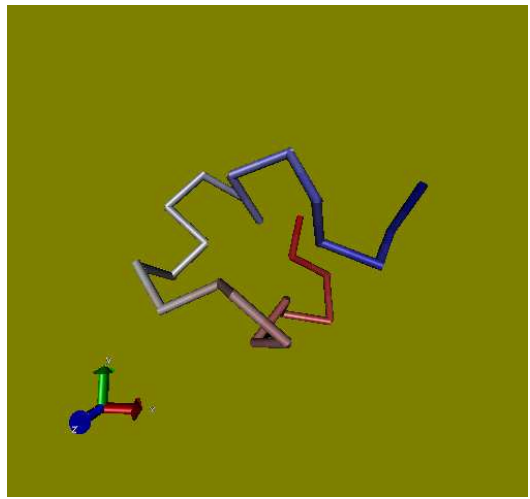}
  \caption{Folded helical structure for model B with 8 bonds.}
  \label{fig:foldedHelix}
\end{figure}

To characterize the (free) energy landscape at a specific temperature, the most common structures are identified and their relative free energies computed at that temperature. Two structures are close in the landscape if they have similar configurations, which means that they have a large number of bonds in common. For model A, the dominant structures are shown in Table~\ref{tab:Table170A}, while those for model B are given in Table~\ref{tab:modelB-dominant}, both for a chain length of $25$.  The dominant structures at low temperatures are designed to be
helical in nature, with long chains allowing for a primitive tertiary structure
in which the helix folds back on itself (see Fig.~\ref{fig:foldedHelix}).
The most common structures at any temperature are those with the lowest Helmholtz free energy at that temperature. Therefore, at low enough temperatures, when the effect of entropy is small, the most common structure is the one with the lowest possible potential energy, which will only have attractive bonds and no repulsive bonds. 
Therefore, unless otherwise specified, here the term ``bond''  refers only to an attractive bond (or hydrogen bond) and not repulsive or covalent bonds.
Using their interaction matrices, it is relatively easy to count the number of occurrences of the different structures and to find the most common structures.

According to the diagram in Fig.~\ref{fig:interactions}(b),  for model B, the maximum number of attractive bonds is 8 for the 25-bead chain. As expected, the most common structure for model B at low temperatures, $\beta^* \geq 4.5$, has 8 attractive bonds (cf. Table~\ref{tab:modelB-dominant}) and therefore has the lowest potential energy for this model. According to Table~\ref{tab:Table170A}, the lowest potential energy configuration in model~A for the 25-bead chain has 21 attractive bonds. However, according to Fig.~\ref{fig:interactions}(a), 36 possible attractive bonds are available for the 25-bead chain in model~A. This means that either the configurations with lower energies that have more than 21 attractive bonds are not geometrically accessible (due to constraints in the model) or their configurational entropies are too low to be observed at these temperatures. It will be shown later (Sec.~\ref{sec:lengtheffects}) that the former scenario is the case. However, if the latter scenario were true, the lower energy configurations would become dominant by reaching lower temperatures.

Within the framework of the model, a folding funnel is identified as 
a region of phase space points corresponding to a set of configurations from which the folded structure is easily and rapidly accessible as the temperature
is lowered. This means that the barriers between local minima located inside the funnel, such as those arising from entropic decreases associated with the formation of new bonds, are small. 
As the temperature is lowered, new minima appear in the funnel region of 
the energy landscape,
corresponding to nearby configurations
that differ in relatively few bonds from the previously favored structures. If barriers 
between nearby states in the landscape are small, the system rapidly equilibrates to the presence of new minima and adopts a more folded
structure.  Although the specific pathway through which the system folds
may involve a number of intermediate structures, the intermediate structures
emerge smoothly with temperature and provide a channel to the folded
structure.  

A quantitative measure of the folding funnel can be obtained by examining
how the dominance $f^*$
of the most preferred structure changes with temperature, where $f^*$ is
the probability of observing the most common configuration.
For real protein systems in which a single, folded structure is thermodynamically stable, one expects that $f^*$ is near unity
for temperatures at which the protein is folded.  Furthermore,
if the protein folds readily as the inverse temperature $\beta$ increases, then
we expect $df^*/d\beta$ to be large and positive in the vicinity of the
inverse folding temperature.

As can be seen in Table~\ref{tab:Table170A}, by decreasing the temperature for model A, some dominant structures are observed, but by decreasing the temperature further, the ratios of their populations to the total population starts to decrease and new structures become dominant. It can be concluded that in this model, the shape of the landscape changes significantly by varying the temperature, where at high temperatures the landscape is riddled with many local minima (many equally preferred structures) and one very deep but wide minimum (no bonded structure), and at low temperatures there are a few narrow deep minima. For model A, either there are deep local minima inside a funnel shaped valley or there are only a few deep local minima beside each other. At the studied temperatures, there is no structure with a very large population, which confirms that there is deep global minimum in the free energy landscape. Since the most common structures at each temperature differ from each other in a few bonds, these deep minima are located close to each other in the landscape but not inside a funnel in the sense that
they are not structures that are adjacent in the configurational space and 
can only be converted into one another through intermediates. 
The barriers involved in these conversions are high enough to make this
a slow process.
For example, as can be seen in Table~\ref{tab:Table170A}, the first two most common structures at $\beta^*=57.5$ differ in seven bonds. Hence there are many barriers that must be overcome to go between the two configurations because seven bonds must be broken and seven new bonds must be formed. On the other hand these two structures share thirteen bonds ($65\%$ of their total bonds), which indicates that they are similar and therefore their locations in the landscape are still relatively close to each other.

\begin{table}[b] 
  \begin{tabular}{r|l|l}
    Rank & most common structure & $f_{\rm obs}$(\%)  \\\hline
    1 & BF BR BV FJ FV JN NR RV &  76.0 $\pm$ 1.2 \\
    2 & BF BR BV FJ JN NR RV & 6.8 $\pm$ 0.8 \\
    3 & BF BV FJ FV JN NR RV & 3.8 $\pm$ 0.6 \\
    4 & BF BR BV FV JN NR RV & 1.9 $\pm$ 0.4 \\
    5 & BF BR BV FJ FV JN RV & 1.3 $\pm$ 0.4 \\
    6 & BF BR BV FJ FV NR RV & 1.0 $\pm$ 0.3 \\
    7 & BF BR BV FJ FV JN NR & 1.0 $\pm$ 0.3 \\
  \end{tabular}
  \caption{Most common configurations of the model~B 25-bead chain at $\beta^*=6$.}
  \label{tab:modelB-dominant-beta6}
\end{table}

Unlike the behavior observed in model A, a single dominant structure is identified in model B by decreasing the temperature, where the probability 
$f^*$ of the most common structure attains a value of nearly one at low temperatures (See Table~\ref{tab:modelB-dominant}). For $\beta^*\geq5.3$, the free energy landscape consists of a single channel in which there are several minima. The most common structures for $\beta^*=6$ are presented in Table~\ref{tab:modelB-dominant-beta6}. 
None of the seven most common structures have a repulsive bond.  This is not surprising, since the formation of a repulsive bond both limits the number of accessible conformations and is energetically unfavorable. The most common structure for $\beta^* \geq 5.3$, BF BR BV FJ FV JN NR RV, is the deepest point in the funnel, and the six other most common structures listed in Table~\ref{tab:modelB-dominant-beta6} differ only in one bond from this structure. This means that there is a funnel-shaped valley with a global minimum corresponding to a folded helix and there are a few local minima of higher free energy beside this deepest point of the landscape. According to Table~\ref{tab:modelB-dominant}, by lowering the temperature the deepest point of the funnel becomes deeper while the other minima become shallower, since the population of the most common structure reaches a value higher than $99.9\%$. This implies that the funnel becomes smoother and steeper as the temperature decreases, and the lowest free energy configuration becomes more accessible.

\begin{figure}[t]
  \centering
  \includegraphics[angle=-90,width=\columnwidth]{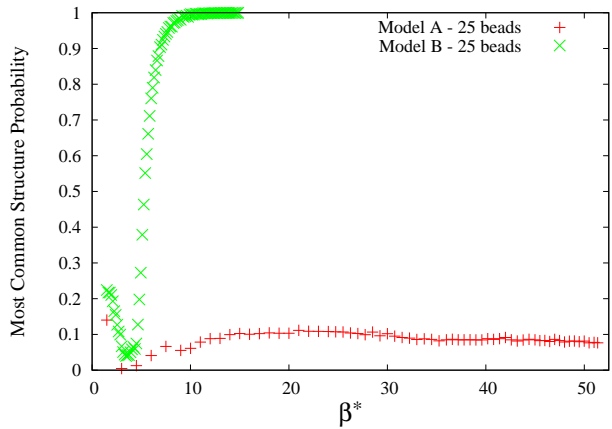}
  \caption{Variation of the probabilities of the most common structure $f^*$ versus the dimensionless inverse temperature $\beta^*$ for chains with 25 beads.}
  \label{fig:funnel}
\end{figure}

The variation of the probability of the most common structure $f^*$ for the two models as a function of temperature is shown in Fig.~\ref{fig:funnel} for chains of 25 beads. For both models, there is a cusp-shaped minimum at which a low-energy structure becomes dominant. The value of the probabilty is very low at the minimum, indicating many competing structures (see Sec.~\ref{sec:threestate}).  For model~A, the probability of the most common structure at low temperatures is fluctuating around a small value of about $0.08$, whereas for model~B, the probability of the most common structure nearly attains unity. This demonstrates once more that the free energy landscape for model A does not have a funnel-like shape.

It will become clear in Sec.~\ref{sec:lengtheffects} that even for model~B, the funnel-like character of the free energy landscape does not persist for chains longer than 29 beads due to geometric frustration.

\subsection{Entropy and free energy calculation for the 25-bead chain in model B}
\label{sec:entropy}

\begin{table}[b] 
  \begin{tabular} {r|l|r|r}
     & configuration & $U_c/\epsilon$ & $S_c/k_B$ \\
    \hline
    1 & AD & 0.9 & 31.3 $\pm$0.8 \\
    2 & No Bond& 0.00 & 31.8 $\pm$0.6 \\
    3 & BF & -1 & 28.6 $\pm$ 0.6\\
    4 & BF JN & -2 & 25.1 $\pm$ 0.6 \\
    5 & BF NR  & -2 & 25.2 $\pm$ 0.6\\
    6 & BF JN RV & -3 & 21.7 $\pm$ 0.4\\
    7 & BF FJ NR RV  & -4 & 17.8 $\pm$ 0.6\\
    8 & BF FJ JN RV  & -4 & 17.6 $\pm$ 0.6\\
    9 & BF FJ JN NR RV & -5 & 13.2 $\pm$ 0.6\\
    10 & BF BR BV FJ JN NR RV & -7 & 3.7 $\pm$ 0.8\\
    11 & BF BV FJ FV JN NR RV & -7 & 2.9 $\pm$ 0.6\\
    12 & BF BR BV FJ FV JN NR RV & -8 & 0
  \end{tabular}
  \caption{Potential energy in units of $\epsilon$ and relative entropy of the most common structures of the model~B 25-bead chain.}
  \label{tab:modelB-dominant-entropy}
\end{table}

As discussed in Sec.~\ref{sec:Entropies}, and as expressed in
Eq.~\eqref{eq:entropyII}, the relative configurational entropies and
consequently the free energy difference of two configurations can be
obtained from the ratio of their probabilities at a specific
temperature.  Since there are fewer possible structures in model~B
than in model~A, the statistical uncertainty in the populations, and
therefore also in the entropies and free energies, is smaller for
model~B. For this reason, and because it is already clear that model~A
does not have a funnel-like free energy landspace, subsequent analysis 
will focus on the characteristics of model~B.

The value of the entropy of a configuration should depend largely on the number of bonds that it has, since the formation of a  bond restricts the distance 
between a specific pair of beads. As can be seen in
Table~\ref{tab:modelB-dominant-entropy}, although the entropies of
configurations with the same number of bonds differ slightly, they are
similar in magnitude.  Typically, the entropy decreases by increasing
the number of bonds due to additional geometric constraints, with the entropy loss typically on the order of $3 k_B$ per bond. Nonetheless,
one sees that configurations with the same energy of -6$\epsilon$ have somewhat different populations and therefore different entropies.

\begin{table}[b] 
  \begin{tabular}{r|l|r|r|r}
    $\beta^* $& configuration & $p_{\rm pred}$ &
    $f_{\rm obs}$ & $\Delta$(\%)\\
    \hline
    1.5& No Bond & 0.206 & 0.165 & 25\\
    1.5 & BF & 0.068 & 0.059 & 15 \\
    1.5 & RV & 0.059 & 0.065 & 9\\
    5.0 & BF BR BV FJ FV JN NR RV & 0.949 & 0.941 & 0.8 \\
    5.0 & BF BR BV FJ JN NR RV& 0.020 & 0.019 & 5\\
    5.0 & BF BV FJ FV JN NR RV & 0.010 & 0.012 & 17\\
    6.0 & BF BR BV FJ FV JN NR RV & 0.988 & 0.980 & 0.8 \\
    6.0 & BF BR BV FJ JN NR RV & 0.005 & 0.005 & 0 \\
    9.0 & BF BR BV FJ FV JN NR RV & 0.999 & 0.999 & 0
  \end{tabular}
  \caption{Comparison of the predicted probability ($p_{\rm pred}$) and the simulation results for the frequency ($f_{\rm obs}$), and their relative difference ($\Delta$), for the most common structures of the model~B 25-bead.}
  \label{tab:Probability-Prediction}
\end{table}

Although in principle, the entropy difference between any two
configurations can be calculated based on the ratio of their
populations, often there is little overlap between the population
distributions of the most common structure at very low temperatures
and the most common structure at very high temperatures (e.g.,
configurations 2 and 12 of Table~\ref{tab:modelB-dominant-entropy}).
Since the configurational entropy difference is independent of
temperature, this difficulty is easily overcome by using one or two
intermediate configurations whose population distribution do have
sufficient overlap at some range of temperatures.  Using the
calculated entropies, one can compute the relative Helmholtz free
energy between any pair of configurations at any temperature. This
allows one to predict the population of any structure at any
temperature and predict the temperature at which the population of two
specific configurations becomes equal.  The free energy and entropy of
some of the most common structures of model~B for the 25-bead chain
are shown in Table~\ref{tab:modelB-dominant-entropy}.

\begin{figure}
  \centering
  \includegraphics[angle=-90,width=\columnwidth]{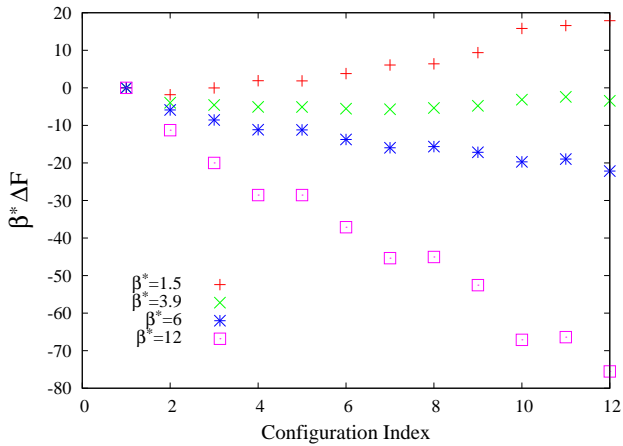}
  \caption{Variation of $\beta^* \Delta F$ versus the configuration index of Table~\ref{tab:modelB-dominant-entropy} for model B, where
$\Delta F$ is the Helmholtz free energy difference with configuration~1 in unit of $\epsilon$.}
  \label{fig:freeenergies}
\end{figure}

The variation of $\beta^* \Delta F$ versus the configuration index of Table~\ref{tab:modelB-dominant-entropy} --- which one could view as a simple way to plot the free energy landscape --- is shown in Fig.~\ref{fig:freeenergies}.  The zero-point of this free energy plot was (arbitrarily) chosen to be the free energy of configuration 1 (AD), i.e., free energies were computed as $\Delta F_{1c}=U_c-TS_c-(U_1-TS_1)$, where $U_c$ and $S_c$ were taken from Table~\ref{tab:modelB-dominant-entropy}. Since both the entropy and the energy of the configurations are decreasing from configuration 1 to 12, the behavior of $\beta^* \Delta F$ is very different for high and low temperatures.
At high temperatures ($\beta^* \le 3$), entropy effects dominate, and the configuration with the largest entropy in Table~\ref{tab:modelB-dominant-entropy} (configuration 2) is the lowest free energy structure for $\beta^*=1.5$ in Fig.~\ref{fig:freeenergies}.  Note that the free energy of other structures increases with increasing number of attractive bonds. In constrast, at lower temperatures, energy effects dominate the free energy landscape, and indeed, in Fig.~\ref{fig:freeenergies}, the configuration from Table~\ref{tab:modelB-dominant-entropy}, which has the lowest potential energy, is seen to be the lowest free energy structure for $\beta^*=6$ and $\beta^*=12$.

\begin{figure}[t] 
  \centering
  \includegraphics[angle=-90,width=\columnwidth]{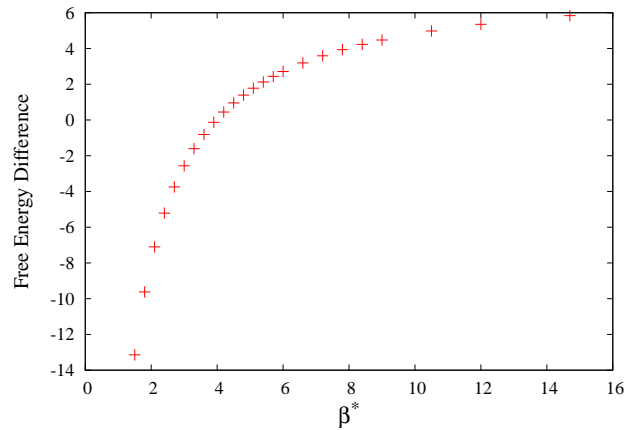}
  \caption{Temperature dependence of the free energy difference of configurations 2 and 12 of 25-bead model B, as listed in Table~\ref{tab:modelB-dominant-entropy}, in units of $\epsilon$.}
  \label{fig:Helmholtzdiff}
\end{figure}

Using the values in Table~\ref{tab:modelB-dominant-entropy}, one can determine that for $\beta^* \geq 4.5$,  the folded helix configuration 
(configuration 12) becomes dominant, since for all the temperatures in that range, this configuration has the lowest free energy. The simulation results for  the population of each configuration confirm this prediction. When configurations are ranked according to their populations, configuration 12 ranks 30th, 13th and 5th for $\beta^*$ values of 4.05, 4.2 and 4.35 respectively, while for $\beta^*\geq4.5$, it ranks first place.

The relative free energy of configurations 2 and 12 is plotted against  $\beta^*$ in Fig.~\ref{fig:Helmholtzdiff}. It can be seen that at $\beta^*\approx4$ their free energies are equal, which implies that their populations are the same. Indeed, simulation results indicate that the populations of configuration 2 and configuration 12 at $\beta^*=3.9$ are 1.0\%\ and 0.5\%, respectively and at $\beta^*=4.05$ are 0.6\%\ and 1.2\%, respectively, which confirms that their population should became equal in the range $3.9  \le \beta^* \le 4.05$.

It should be noted that in our calculations,
structures that have a population of less than 
0.5\%\ 
are not considered to simplify computations.
As a result, when calculating the probabilities of the configurations with 25 beads, only $78$ configurations were used. Although this  introduces a systematic error,
the predicted probabilities are very close to the observed ones from the simulation runs, as can be seen in Table~\ref{tab:Probability-Prediction}. According to this table, the predicted values agree better with the simulation results at lower temperatures. The disagreement is due to the fact that some configurations with very low populations have not been considered in the probability calculations, but since these configuration occur more frequently at high temperatures, neglecting their contribution leads to a larger error at high temperatures.

\subsection{Entropy and free energy calculation for the 35-bead model B chain}

The entropies and free energies of 35-bead configurations are calculated in a similar way to the 25-bead case. Adding only 10 beads to the chain changes the number of possible attractive bonds from 8 in the 25-bead chain to 23 in the 35-bead chain (cf.~\ref{fig:interactions}), which results in a much more complex energy landscape. 

The dramatic change in landscape can be seen in
Table~\ref{tab:modelB-dominant35} and Fig.~\ref{fig:29-35}, where we
see that, unlike the 25-bead chain, the probability of the most common
structure does not approach unity even at very low temperatures.

As can be seen in Table~\ref{tab:modelB-dominant35}, by increasing $\beta^*$ (decreasing temperature) a few structures become dominant at different temperatures. Except for the lowest energy configuration with 23 attractive bonds, other energies are degenerate with multiple configurations possessing the same number of bonds. It will be shown in the next section that a structure with 23 attractive bonds is geometrically prohibited. In fact, configurations with 21, 22, or 23 attractive bonds have not been observed in any simulation runs.

\begin{table*} 
  \begin{tabular}{r|l|r}
    $\beta^*$ & the most common structure & $f_{\rm obs}$(\%)  \\
    \hline
    1.5& No Bond & 11.7 $\pm$ 1.3 \\
    5.25  & BF BZ Bd Bh FJ FV FZ Fd Fh JN Jd Jh NR Nd RV VZ Zd dh  & 7.2 $\pm$ 0.9 \\
    9.0 & BF BZ Bd Bh FJ FV FZ Fd Fh JN Jd Jh NR Nd RV VZ Zd dh &  18.8 $\pm$ 1.5 \\
    16.5 & BF BR BV BZ Bh FJ FZ Fd Fh JN Jd Jh NR Nh RV Rh VZ Zd dh & 25.8 $\pm$ 1.8 \\
    31.5  & BF BR BV BZ Bh FJ FZ Fd Fh JN Jd Jh NR Nh RV Rh VZ Zd dh & 24.7 $\pm$ 1.6 \\
    53.63 & BF BR BV BZ Bh FJ FZ Fd Fh JN Jd Jh NR Nh RV Rh VZ Zd dh & 23.6 $\pm$ 1.6 \\
    \hline
     $\beta^*$ & the second most common structure & $f_{\rm obs}$(\%)  \\
    \hline
    1.5 & dh & 2.3 $\pm$ 0.6\\
    5.25 & BF BR BV BZ Bd Bh FJ Fd Fh JN Jh NR Nh RV Rh VZ Zd dh  & 5.3 $\pm$ 0.9 \\
    9.0 & BF BR BV BZ Bh FJ FZ Fd Fh JN Jd Jh NR Nh RV Rh VZ Zd dh & 15.4 $\pm$ 1.5  \\
    16.5 &  BF BZ Bd Bh FJ FV FZ Fd Fh JN Jd Jh NR Nd Nh RV VZ Zd dh & 14.4 $\pm$ 1.4 \\
    31.5 & BF BZ Bd Bh FJ FV FZ Fd Fh JN Jd Jh NR Nd Nh RV VZ Zd dh & 16.7 $\pm$ 1.4 \\
    53.63 & BF BZ Bd Bh FJ FV FZ Fd Fh JN Jd Jh NR Nd Nh RV VZ Zd dh & 15.2 $\pm$ 1.3
  \end{tabular}
  \caption{Most common configurations of the model~B 35-bead chain.}
  \label{tab:modelB-dominant35}
\end{table*}

One difference between the landscape of the 35-bead chain and that of the 25-bead chain is the magnitude of the entropic barriers between configurations with different energies. The most common structures in Table~\ref{tab:modelB-dominant35} at high $\beta^*$ have an energy of $-19\epsilon$. Beside the two main configurations with the energy of $-19\epsilon$, which are presented in Table~\ref{tab:modelB-dominant35}, there are at least 18 other configurations with the same potential energy but with lower entropies.  Three structures with an energy of $-20 \epsilon$ and with relatively low entropies have been observed in the runs as well, but,  according to Table~\ref{tab:modelB-dominant35}, these were never among the first two most common structures.  The configuration with the potential energy of $-20 \epsilon$ that has the highest entropy is different in five bonds from the most common configuration of Table~\ref{tab:modelB-dominant35}.
This implies that there is a substantial entropic barrier between these configurations.

A second difference with the 25-bead case is that a few different configurations of the 35-bead chain exist at low temperatures and are observed with nearly the same frequency.   For example, as Table~\ref{tab:modelB-dominant35} shows, the two most common structures for $16.5 \le \beta^* \le 53$ have 19 attractive bonds. While these two structures differ slightly in their populations, structurally they differ by more than one bond, quite unlike the seven most common structures of the 25-bead chain at $\beta^*=6$ (cf. Fig.~\ref{tab:modelB-dominant-beta6}) which only differ from each other by one bond. Since the most common structures of the 35-bead chain at low temperatures share most of their bonds, they are near one another in the energy landscape.  However, since the most common structure differs from other common structures by more than one bond, they do not necessarily lie inside a single valley in the landscape. A more plausible interpretation is that the landscape at low temperatures for the 35-bead chain consists of several minima that are close but not necessarily inside the same channel, and that the landscape does not have a single deep minumum at very low temperatures.

A final difference with the 25-bead case that becomes apparent is that the range of energies and that of entropies for the observed configurations are $8 \epsilon$ and $32k_B$ respectively for 25-bead chains, while these 
are  $20\epsilon$ and $140k_B$ respectively for 35-bead chains. This confirms the view that the landscape of the 35-bead chain is much wider than the 25-bead chain landscape. This also shows that studying the landscape requires a much wider range of temperatures and more replicas.

\subsection{Effects of the protein-like chain length}
\label{sec:lengtheffects}

\begin{figure}[t]
  \centering
  \includegraphics[angle=-90,width=\columnwidth]{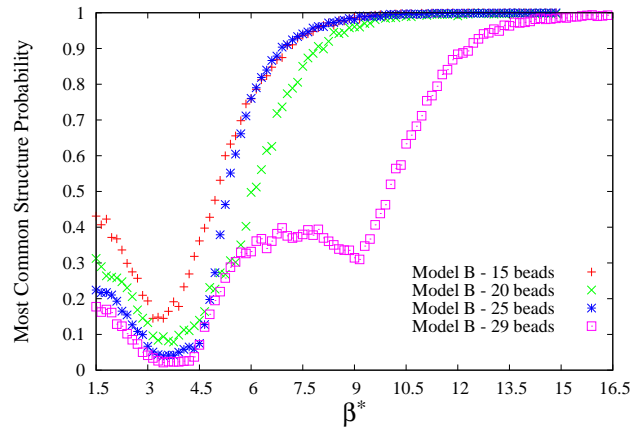}
  \caption{Variation of the probabilities of the most common structure, $f^*$, versus the $\beta^*$ for chains with 15, 20, 25 and 29 beads}
  \label{fig:15-29}
\end{figure}

\begin{figure}[t]
  \centering
  \includegraphics[angle=-90,width=\columnwidth]{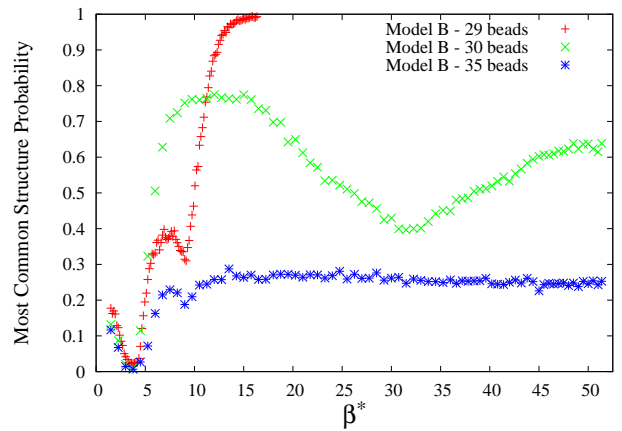}
  \caption{Variation of the probabilities of the most common structure, $f^*$, versus the $\beta^*$ for the chains with 29, 30 and 35 beads. The result of the 29-bead chain from figure~\ref{fig:15-29} is presented here as a reference.}
  \label{fig:29-35}
\end{figure}

For 25-bead chains, the probability of the most common structure
approaches unity at low temperatures, while the longer 35-bead chain
did not show this trend. There are two possible reasons for this
behavior. First, it is possible that the studied range of temperatures
was not sufficiently large to observe the lowest energy configuration
for long chains in the simulation. The second possible reason is
that the lowest possible energy is not geometrically accessible
considering the criteria of model B.  The effect of the inaccessibility of the lowest
energy configuration is that
several structures with the same energy compete for the highest
probability. While the configurational entropies of these
structures are somewhat different, there is no configuration with a much
higher entropy than all the other structures with the same energy, and
hence none of their maximum structural probabilities approaches unity in the
accessible temperature range. It turns out that the second scenario is much more plausible.  To understand why, it is helpful to consider the thermodynamic characteristics of model B for other chain lengths.
For chains of length 15, 20, 25, 29, 30 and 35, the maximum number of attractive bonds are 3, 5, 8, 12, 17 and 23, respectively. The temperature dependence of the probability of the most common structure for these cases is shown in Figs.~\ref{fig:15-29} and~\ref{fig:29-35}. 

One sees in Fig.~\ref{fig:15-29} that for chains with 15, 20, 25 and 29 beads, after going through one or two minima, the probability of the most common structure $f^*$  approaches unity at low temperatures. In these cases, the most probable configurations are also the ones with the lowest energy, i.e., with the maximum number of attractive bonds.  
For the 29-bead chain there is a distinctive peak in the probability of the most common structure at $\beta^*\approx7.5$, which can be explained by the large entropy difference between the most common structure with 11 bonds and the most common structure with 12 bonds, which allows the 11-bond configuration to become the most common structure for $4.35 \le \beta^* \le 9$.
Apparently, at $\beta^*=9$, the energy difference becomes equal to the entropy difference times $T^*$, so that for $\beta^* > 9$ the structure with 12 bonds becomes the most common structure.

Fig.~\ref{fig:29-35} shows that the situation is quite different for longer chains. 
For the 30-bead chain, the maximum possible number of bonds is 17, but no such structure was observed in the simulations, even when using different numbers of replicas, different PT temperature sets and different ranges of temperatures. This strongly suggests that that it is impossible to satisfy the geometric constraints needed to form all possible bonds.
Once the geometric constraints cannot all be satisfied for one particular chain length, this automatically implies that they can also not be satisfied for longer chains. Indeed, in the 35 bead case, the lowest energy configuration is also not observed.

As can be seen in Fig.~\ref{fig:29-35}, when $4.5\le \beta^*\le 7.5$, the probability of the most common structure increases for the 30-bead chain (similar to the behavior observed in 15, 20, 25 and 29 beads chain systems). The probability of the most common structure then remains more or less unchanged up to  $\beta^*\approx15$. After this plateau region, the probability decreases until reaching a $\beta^*$ value at which the probability of the two most common structures becomes equal (in this case, these are the 15-bond structure with the highest entropy and the 16-bond structure with the highest entropy), which can be seen as a  minimum in the graph. After passing this local minimum, the structure with 16 bonds becomes the most common structure. However, because there are at least six structures with 16 bonds, the probability of the most common structure is not close to one even at very low temperatures. One explanation is that the structure with 17 bonds is geometrically prohibited, leading to several energetically degenerate configurations with 16 bonds to become common at low temperatures (their relative populations depending on configurational entropy differences).

\begin{figure} 
  \centering
  \includegraphics[angle=-90,width=\columnwidth]{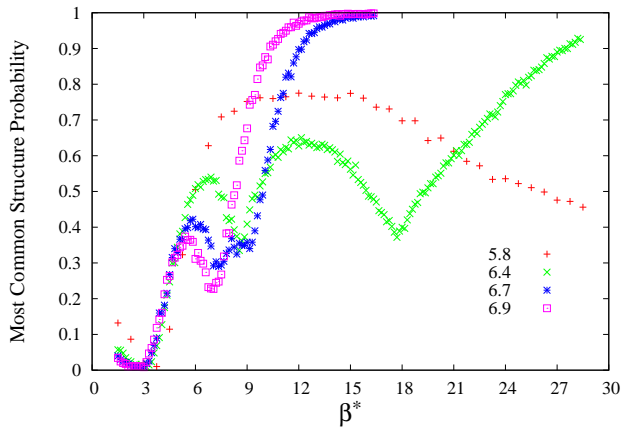}
  \caption{Variation of the probabilities of the most common structure versus the $\beta^*$ for the 30-bead chain for different attractive bond interaction distances (increasing $\sigma_2$ from the initial 5.8 \AA\ to 6.4 \AA\, 6.7 \AA\ and 6.9 \AA\ }
  \label{fig:differentranges}
\end{figure}

A second argument for the geometric frustration of the lowest energy
configuration for larger chain lengths can be found by slowly
relaxing the geometric restrictions imposed by the range of interaction
of the attractive bonds in the model. If the configuration is
geometrically prohibited, then by slowly increasing the bonding distance one should
find a critical value of the range at which the configuration suddenly
becomes accessible, and since its energy is lower than any other
structure, that configuration should at the same time suddenly become
a very common, if not the most common, structure.

The attractive bonds can be formed at a range 4.6\AA\ $ \le r_{ij} \le$5.8\AA\ ($\sigma_1$=4.6 \AA\ and $\sigma_2$=5.8 \AA), where $r_{ij}$ is the distance between beads $i$ and $j$.
To change the attractive range, only $\sigma_2$ was increased. At $\sigma_2=6.2$\AA, it was possible to observe the lowest energy configuration for the 30 bead chain (with $17$ attractive bonds) at low temperatures, while this structure was not observed for the runs with $\sigma_2 \le 6.1$ \AA. Figure.~\ref{fig:differentranges} illustrates this by plotting the probability of the most common structure as a function of temperatures for several values of $\sigma_2$. For $\sigma_2=6.4$~\AA, the probability of the 17 bonds structure approaches one around $\beta^*=27$, and by increasing the value of $\sigma_2$, this occurs at lower $\beta^*$, since the entropic barriers between the low energies configurations, such as the configurations with 15 bonds and 16 bonds, become smaller. The first bump in Fig.~\ref{fig:differentranges} represents a temperature region where the structure with $15$ bonds becomes the most probable configuration, and the second bump occurs at higher $\beta^*$ values, where a $16$-bond configuration becomes the most probable structure. Since the entropy difference between the configurations with different energies becomes smaller for larger $\sigma_2$, this range of $\beta^*$, where the structure with 15 bonds becomes the most common structure, becomes smaller for larger $\sigma_2$ values as can be seen for $\sigma_2=$6.7 \AA\ and $\sigma_2=$6.9 \AA\ in Fig.~\ref{fig:differentranges}.

We conclude that for chains smaller than 30 beads 
the landscape consists of one deep funnel at low temperatures that contains several
minima. The funnel becomes steeper by decreasing the temperature. At very low temperatures the landscape consists of a
smooth funnel with a very deep global minimum representing the configuration with the
maximum number of bonds. But for chains longer than 29 beads, the
landscape of the longer chains does not consist of one deep funnel,
even for low temperatures.  Rather, it consists of several minima
or channels between which there are entropic barriers that increase with
increasing chain length.

\section{Conclusions}
\label{sec:concl}

In this work two different models of a protein-like chain that differ primarily in the number of attractive interactions were introduced 
and the characteristics of their free energy landscape was analyzed.  Fewer bonding interactions are present in the second model (model B), leading to a system with less frustration and a free energy landscape that possesses fewer local minima.
The models were designed to encourage the formation of helical secondary structural elements and such helices were observed in model~B at low temperatures. For long enough chains ($>17$ beads), model B also allows a tertiary structure.

It was shown that for model B, the free energy landscape of the 25-bead chain has a smooth funnel that has important effects on both the dynamics and the thermodynamics of the system. In this model, the free energy landscape at low temperatures contains a deep valley with several minima around it located inside one basin. As the temperature decreases, the deepest point of the funnel becomes deeper, while the minima around the deepest point become shallower. This trend continues until a temperature is reached in which all local minima in the free energy landscape have vanished and only a single global minimum exists. In contrast to Model~B, Model~A 
does not exhibit a preference for a specific native structure at low temperatures. This may be attributed to several factors, such as the lack of rigidity of the chain in this model, several large entropic barriers, and the possibility of having many structures with the same energy.

It was shown that the relative configurational entropy is temperature independent. Hence, using the populations of the configurations at different temperatures, the relative free energy and entropy of any pair of configurations can be calculated. From the free energies of different structures at the studied temperatures, the populations of all configurations at any temperature were predicted and verified against simulation results. These results agree reasonably with the simulation results, which shows one of the great advantages of using discontinuous potentials to study the free energy landscape.

In model B, the single funnel morphology of the free energy landscape
persists for chains up to 29 beads long.  However, for chains
of 30 beads or longer, the simulation results strongly suggest
that the structure satisfying all possible attractive bonds is
geometrically prohibited, while at the same time, the entropic
barriers between the configurations with different energies become
larger.  For long chains, 
the landscape at low temperatures 
consists of a few distinct channels that are relatively close to each other but separated by high barriers.

The observed landscape can provide insight into the shape of the landscape of actual proteins. While for small chains the native structure seems to be the lowest free energy structure, the existence of several distinct funnels in the landscape of long chains suggests the possibility that the native structure of real proteins is not necessarily the lowest free energy structure but may correspond to a configurational basin that can be accessed easily during the folding dynamics. 
Another factor that should be considered for long proteins is the important effect of temperature on the morphology of the landscape. In our study, the basin containing the global minimum becomes steeper as the temperature decreases for short chains. However, for longer chains, the basin becomes steeper while the deepest point of the landscape can shift from one configuration to another configuration with slightly different bonds over the same temperature range. Thus, for long proteins, the structure may be more sensitive to temperature fluctuations and by slightly changing the temperature the thermodynamically stable configuration can shift to a configuration that differs substantially.

The simulation results presented here can be used to analyze
the dynamics of the protein-like chain
by computing the first-passage time solution for the transition rates among the
individual microstates.  
The individual rates between microstates can then be
incorporated into a Markovian model of the relaxation of the chain
and the dynamics of the folding process examined to probe how features in the energy landscape
determine the relaxation profile of the protein-like chain\cite{chainDynamics}.

\begin{acknowledgments}

Computations were performed on the GPC supercomputer at the SciNet HPC Consortium, which is funded by the Canada Foundation for Innovation under the auspices of Compute Canada, the Government of Ontario, the Ontario Research Fund Research Excellence and the University of Toronto.

This work was supported by a grant from the Natural Sciences and Engineering Research Council of Canada.

\end{acknowledgments}


\begin{thebibliography}{27}
\expandafter\ifx\csname natexlab\endcsname\relax\def\natexlab#1{#1}\fi
\expandafter\ifx\csname bibnamefont\endcsname\relax
  \def\bibnamefont#1{#1}\fi
\expandafter\ifx\csname bibfnamefont\endcsname\relax
  \def\bibfnamefont#1{#1}\fi
\expandafter\ifx\csname citenamefont\endcsname\relax
  \def\citenamefont#1{#1}\fi
\expandafter\ifx\csname url\endcsname\relax
  \def\url#1{\texttt{#1}}\fi
\expandafter\ifx\csname urlprefix\endcsname\relax\def\urlprefix{URL }\fi
\providecommand{\bibinfo}[2]{#2}
\providecommand{\eprint}[2][]{\url{#2}}

\bibitem[{\citenamefont{Dill et~al.}(2007)\citenamefont{Dill, Ozkan, Weikl,
  Chodera, and Voelz}}]{Dill:64}
\bibinfo{author}{\bibfnamefont{K.~A.} \bibnamefont{Dill}},
  \bibinfo{author}{\bibfnamefont{S.~B.} \bibnamefont{Ozkan}},
  \bibinfo{author}{\bibfnamefont{T.~R.} \bibnamefont{Weikl}},
  \bibinfo{author}{\bibfnamefont{J.~D.} \bibnamefont{Chodera}},
  \bibnamefont{and} \bibinfo{author}{\bibfnamefont{V.~A.} \bibnamefont{Voelz}},
  \bibinfo{journal}{Current Opinion in Structural Biology}
  \textbf{\bibinfo{volume}{17}}, \bibinfo{pages}{342} (\bibinfo{year}{2007}).

\bibitem[{\citenamefont{Leopold et~al.}(1992)\citenamefont{Leopold, Montal, and
  Onuchic}}]{Onuchic:65}
\bibinfo{author}{\bibfnamefont{P.~E.} \bibnamefont{Leopold}},
  \bibinfo{author}{\bibfnamefont{M.}~\bibnamefont{Montal}}, \bibnamefont{and}
  \bibinfo{author}{\bibfnamefont{J.~N.} \bibnamefont{Onuchic}},
  \bibinfo{journal}{Proc. Natl. Acad. Sci. USA} \textbf{\bibinfo{volume}{89}},
  \bibinfo{pages}{8721} (\bibinfo{year}{1992}).

\bibitem[{\citenamefont{Dill}(1993)}]{Dill:62}
\bibinfo{author}{\bibfnamefont{K.~A.} \bibnamefont{Dill}},
  \bibinfo{journal}{Curr. Opinion Struct. Biol.} \textbf{\bibinfo{volume}{3}},
  \bibinfo{pages}{99} (\bibinfo{year}{1993}).

\bibitem[{\citenamefont{Bryngelson et~al.}(1995)\citenamefont{Bryngelson,
  Onuchic, Socci, and Wolynes}}]{Wolynes:45}
\bibinfo{author}{\bibfnamefont{J.~D.} \bibnamefont{Bryngelson}},
  \bibinfo{author}{\bibfnamefont{J.~N.} \bibnamefont{Onuchic}},
  \bibinfo{author}{\bibfnamefont{N.~D.} \bibnamefont{Socci}}, \bibnamefont{and}
  \bibinfo{author}{\bibfnamefont{P.~G.} \bibnamefont{Wolynes}},
  \bibinfo{journal}{Proteins:Struct. Funct. Genet.}
  \textbf{\bibinfo{volume}{21}}, \bibinfo{pages}{167}
  (\bibinfo{year}{1995}).

\bibitem[{\citenamefont{Wolynes et~al.}(1995)\citenamefont{Wolynes, Onuchi, and
  Thirumulai}}]{Thirumulai:67}
\bibinfo{author}{\bibfnamefont{P.~G.} \bibnamefont{Wolynes}},
  \bibinfo{author}{\bibfnamefont{J.~N.} \bibnamefont{Onuchi}},
  \bibnamefont{and}
  \bibinfo{author}{\bibfnamefont{D.}~\bibnamefont{Thirumulai}},
  \bibinfo{journal}{Science} \textbf{\bibinfo{volume}{267}},
  \bibinfo{pages}{1619} (\bibinfo{year}{1995}).

\bibitem[{\citenamefont{Onuchic et~al.}(1995)\citenamefont{Onuchic, Wolynes,
  Luthey-Schulten, and Socci}}]{Socci:20}
\bibinfo{author}{\bibfnamefont{J.~N.} \bibnamefont{Onuchic}},
  \bibinfo{author}{\bibfnamefont{P.~G.} \bibnamefont{Wolynes}},
  \bibinfo{author}{\bibfnamefont{Z.}~\bibnamefont{Luthey-Schulten}},
  \bibnamefont{and} \bibinfo{author}{\bibfnamefont{N.~D.} \bibnamefont{Socci}},
  \bibinfo{journal}{Proc. Natl. Acad. Sci. USA} \textbf{\bibinfo{volume}{92}},
  \bibinfo{pages}{3626} (\bibinfo{year}{1995}).

\bibitem[{\citenamefont{Onuchic et~al.}(1996)\citenamefont{Onuchic, Socci,
  Luthey-Schulten, and Wolynes}}]{Wolynes:21}
\bibinfo{author}{\bibfnamefont{J.~N.} \bibnamefont{Onuchic}},
  \bibinfo{author}{\bibfnamefont{N.~D.} \bibnamefont{Socci}},
  \bibinfo{author}{\bibfnamefont{Z.}~\bibnamefont{Luthey-Schulten}},
  \bibnamefont{and} \bibinfo{author}{\bibfnamefont{P.~G.}
  \bibnamefont{Wolynes}}, \bibinfo{journal}{Fold Des.}
  \textbf{\bibinfo{volume}{1}}, \bibinfo{pages}{441} (\bibinfo{year}{1996}).

\bibitem[{\citenamefont{Dill and Chan}(1997)}]{Dill:66}
\bibinfo{author}{\bibfnamefont{K.~A.} \bibnamefont{Dill}} \bibnamefont{and}
  \bibinfo{author}{\bibfnamefont{H.~S.} \bibnamefont{Chan}},
  \bibinfo{journal}{Nature Structural Biology} \textbf{\bibinfo{volume}{4}},
  \bibinfo{pages}{10} (\bibinfo{year}{1997}).

\bibitem[{\citenamefont{Socci et~al.}(1998)\citenamefont{Socci, Onuchic, and
  Wolynes}}]{Wolynes:22}
\bibinfo{author}{\bibfnamefont{N.~D.} \bibnamefont{Socci}},
  \bibinfo{author}{\bibfnamefont{J.~N.} \bibnamefont{Onuchic}},
  \bibnamefont{and} \bibinfo{author}{\bibfnamefont{P.~G.}
  \bibnamefont{Wolynes}}, \bibinfo{journal}{Proteins}
  \textbf{\bibinfo{volume}{32}}, \bibinfo{pages}{136} (\bibinfo{year}{1998}).

\bibitem[{\citenamefont{Gin et~al.}(2009)\citenamefont{Gin, Garrahan, and
  Geissler}}]{Gin:19}
\bibinfo{author}{\bibfnamefont{B.~C.} \bibnamefont{Gin}},
  \bibinfo{author}{\bibfnamefont{J.~P.} \bibnamefont{Garrahan}},
  \bibnamefont{and} \bibinfo{author}{\bibfnamefont{P.~L.}
  \bibnamefont{Geissler}}, \bibinfo{journal}{J. Mol. Biol.}
  \textbf{\bibinfo{volume}{392}}, \bibinfo{pages}{1303} (\bibinfo{year}{2009}).

\bibitem[{\citenamefont{McLeish}(2005)}]{McLeish:24}
\bibinfo{author}{\bibfnamefont{T.}~\bibnamefont{McLeish}},
  \bibinfo{journal}{Biophys. J.} \textbf{\bibinfo{volume}{88}},
  \bibinfo{pages}{172} (\bibinfo{year}{2005}).

\bibitem[{\citenamefont{Duane et~al.}(1987)\citenamefont{Duane, Kennedy,
  Pendleton, and Roweth}}]{Duane:30}
\bibinfo{author}{\bibfnamefont{S.}~\bibnamefont{Duane}},
  \bibinfo{author}{\bibfnamefont{A.~D.} \bibnamefont{Kennedy}},
  \bibinfo{author}{\bibfnamefont{B.~J.} \bibnamefont{Pendleton}},
  \bibnamefont{and} \bibinfo{author}{\bibfnamefont{D.}~\bibnamefont{Roweth}},
  \bibinfo{journal}{Phys. Lett. B} \textbf{\bibinfo{volume}{195}},
  \bibinfo{pages}{216} (\bibinfo{year}{1987}).

\bibitem[{\citenamefont{Swendsen and Wang}(1986)}]{Swendsen:6}
\bibinfo{author}{\bibfnamefont{R.~H.} \bibnamefont{Swendsen}} \bibnamefont{and}
  \bibinfo{author}{\bibfnamefont{J.-S.} \bibnamefont{Wang}},
  \bibinfo{journal}{Phys. Rev. Lett.} \textbf{\bibinfo{volume}{57}},
  \bibinfo{pages}{2607} (\bibinfo{year}{1986}).

\bibitem[{\citenamefont{Geyer}(1991)}]{Greyer:7}
\bibinfo{author}{\bibfnamefont{C.~J.} \bibnamefont{Geyer}}, in
  \emph{\bibinfo{booktitle}{Proceedings of the 23rd Symposium on the Interface:
  Computing Science and Statistics}} (\bibinfo{year}{1991}), pp.
  \bibinfo{pages}{156}.

\bibitem[{\citenamefont{Tesi et~al.}(1996)\citenamefont{Tesi, van Rensburg,
  Orlandini, and Whittington}}]{whittington:18}
\bibinfo{author}{\bibfnamefont{M.~C.} \bibnamefont{Tesi}},
  \bibinfo{author}{\bibfnamefont{E.~J.~J.} \bibnamefont{van Rensburg}},
  \bibinfo{author}{\bibfnamefont{E.}~\bibnamefont{Orlandini}},
  \bibnamefont{and} \bibinfo{author}{\bibfnamefont{S.~G.}
  \bibnamefont{Whittington}}, \bibinfo{journal}{J. Statist. Phys.}
  \textbf{\bibinfo{volume}{82}}, \bibinfo{pages}{155} (\bibinfo{year}{1996}).

\bibitem[{\citenamefont{Earl and Deem}(2005)}]{Earl:8}
\bibinfo{author}{\bibfnamefont{D.~J.} \bibnamefont{Earl}} \bibnamefont{and}
  \bibinfo{author}{\bibfnamefont{M.~W.} \bibnamefont{Deem}},
  \bibinfo{journal}{Phys. Chem. Chem. Phys.,} \textbf{\bibinfo{volume}{7}},
  \bibinfo{pages}{3910} (\bibinfo{year}{2005}).

\bibitem[{\citenamefont{Athawale et~al.}(2007)\citenamefont{Athawale, Goel,
  Ghosh, Truskett, and Garde}}]{Athawale:13}
\bibinfo{author}{\bibfnamefont{M.~V.} \bibnamefont{Athawale}},
  \bibinfo{author}{\bibfnamefont{G.}~\bibnamefont{Goel}},
  \bibinfo{author}{\bibfnamefont{T.}~\bibnamefont{Ghosh}},
  \bibinfo{author}{\bibfnamefont{T.~M.} \bibnamefont{Truskett}},
  \bibnamefont{and} \bibinfo{author}{\bibfnamefont{S.}~\bibnamefont{Garde}},
  \bibinfo{journal}{Proc. Natl. Acad. Sci. USA} \textbf{\bibinfo{volume}{104}},
  \bibinfo{pages}{733} (\bibinfo{year}{2007}).

\bibitem[{\citenamefont{Bellemans et~al.}(1980)\citenamefont{Bellemans, Orban,
  and van Belle}}]{Bellemans:26}
\bibinfo{author}{\bibfnamefont{A.}~\bibnamefont{Bellemans}},
  \bibinfo{author}{\bibfnamefont{J.}~\bibnamefont{Orban}}, \bibnamefont{and}
  \bibinfo{author}{\bibfnamefont{D.}~\bibnamefont{van Belle}},
  \bibinfo{journal}{Mol. Phys.} \textbf{\bibinfo{volume}{39}},
  \bibinfo{pages}{781} (\bibinfo{year}{1980}).

\bibitem[{\citenamefont{Zhou and Karplus}(1997)}]{ZHOU:15}
\bibinfo{author}{\bibfnamefont{Y.}~\bibnamefont{Zhou}} \bibnamefont{and}
  \bibinfo{author}{\bibfnamefont{M.}~\bibnamefont{Karplus}},
  \bibinfo{journal}{Proc. Natl. Acad. Sci. USA} \textbf{\bibinfo{volume}{94}},
  \bibinfo{pages}{14429} (\bibinfo{year}{1997}).

\bibitem[{\citenamefont{Li and Bru¨schweiler}(2009)}]{Wei:31}
\bibinfo{author}{\bibfnamefont{D.-W.} \bibnamefont{Li}} \bibnamefont{and}
  \bibinfo{author}{\bibfnamefont{R.}~\bibnamefont{Bru¨schweiler}},
  \bibinfo{journal}{Phys. Rev. Lett.} \textbf{\bibinfo{volume}{102}},
  \bibinfo{pages}{118108} (\bibinfo{year}{2009}).

\bibitem[{\citenamefont{Rapaport}(2004)}]{Rappa:4}
\bibinfo{author}{\bibfnamefont{D.~C.} \bibnamefont{Rapaport}},
  \emph{\bibinfo{title}{The art of molecular dynamics simulation}}
  (\bibinfo{publisher}{Cambridge University Press},
  \bibinfo{address}{Cambridge}, \bibinfo{year}{2004}), \bibinfo{edition}{2nd}
  ed.

\bibitem[{\citenamefont{{Hern{\'{a}}ndez de la Pe{\~{n}}a}
  et~al.}(2007)\citenamefont{{Hern{\'{a}}ndez de la Pe{\~{n}}a}, van Zon,
  Schofield, and Opps}}]{Hernandez:9}
\bibinfo{author}{\bibfnamefont{L.}~\bibnamefont{{Hern{\'{a}}ndez de la
  Pe{\~{n}}a}}}, \bibinfo{author}{\bibfnamefont{R.}~\bibnamefont{van Zon}},
  \bibinfo{author}{\bibfnamefont{J.}~\bibnamefont{Schofield}},
  \bibnamefont{and} \bibinfo{author}{\bibfnamefont{S.~B.} \bibnamefont{Opps}},
  \bibinfo{journal}{J. Chem. Phys.} \textbf{\bibinfo{volume}{126}},
  \bibinfo{pages}{074105} (\bibinfo{year}{2007}).

\bibitem[{\citenamefont{Movahed et~al.}(2012)\citenamefont{Movahed, van Zon,
  and Schofield}}]{chainDynamics}
\bibinfo{author}{\bibfnamefont{H.~B.} \bibnamefont{Movahed}},
  \bibinfo{author}{\bibfnamefont{R.}~\bibnamefont{van Zon}}, \bibnamefont{and}
  \bibinfo{author}{\bibfnamefont{J.}~\bibnamefont{Schofield}},
  \bibinfo{journal}{in preparation}  (\bibinfo{year}{2012}).

\end{thebibliography}

\end{document}